\documentclass[aps,pra,amssymb,preprint]{revtex4-1}
\usepackage{graphicx}
\usepackage{rotating}
\usepackage{tabulary}
\usepackage{xcolor}
\usepackage[T1]{fontenc}
\begin{document}

\title{Noise induced Non-Markovianity }

\author{Arzu Kurt}
\author{Resul Eryigit}
\email{arzukurt@ibu.edu.tr}
\affiliation{Department of Physics, Abant Izzet Baysal University, Bolu, Turkey}
\date{\today}

\begin{abstract}
We have studied the non-Markovianity of dichotomously driven spin-boson model in the strong coupling regime in both memory kernel and time convolutionless master equation formulations. A strong correlation between the decay time of the environmental correlation function and the non-zero non-Markovianity is found in the absence of the external noise. Stochastic driving is shown to create strong non-Markovianity when the dynamics of the system without driving is Markovian. Also, exact analytical expressions for the trace distance distinguishability and the non-Markovianity were obtained for the certain range of the parameters that describe the system and its environment.
\end{abstract}
\maketitle
\section{Introduction}

Non-Markovianity of open quantum system dynamics which refers to nontrivial memory effects has been one of the most active areas of research due to both its relevance for quantum information science and possible applications in quantum technologies~\cite{rivas14,breuer16,devega17}.
Various context-dependent measures, such as trace-distance based distinguishability~\cite{breuer09,laine10}, divisibility of dynamical maps~\cite{rivas10},
Fisher information, classically assisted entanglement, violation of quantum regression theorem and linear response have been developed to quantify the non-Markovianity of the system dynamics. Such measures have been used to study memory effects in the dynamics of many quantum systems, such as qubit(s) driven by classical noise~\cite{benedetti13,benedetti14,rossi16},
spin-boson model~\cite{clos12,haikka13,addis14,lofranco14,schmidt16,liu17} and photosynthetic systems~\cite{chen14,liu15,chen15} theoretically. Several experimental realizations of non-Markovianity control in optical~\cite{cialdi17,bernardes15,chiuri12,liu11} and solid state settings~\cite{haase18,peng18,wang18} are, also, reported. Non-Markovianity of spin-boson model in the weak coupling regime was studied in the temperature-cutoff frequency of the environmental spectral function plane by Clos and Breuer~\cite{clos12} who found, by using a time convolutionless master equation approach, that its dynamics are non Markovian for the low temperature and the low cutoff frequency region. Temperature and interaction strength dependence of the non-Markovianity for the spin-boson model were
discussed in Refs.~\cite{liu15,chen15} in the context of photosynthetic systems. Liu et al~\cite{liu15} have studied the non-Markovianity in the chromophore-qubit system dynamics as function of bath temperature and the coupling constant both in weak and strong coupling regimes with polaron master equation for super-Ohmic bath spectral density and found that the increasing temperature leads to reduction in non-Markovianity in both weak and strong interaction cases while increasing coupling constant enhances non-Markovianity in weak coupling and diminishes in the strong coupling case. While increasing temperature was found to enhance $\mathcal{N}$ in the strong coupling regime in Ref.~\cite{chen15}, opposite was found in Ref.~\cite{liu15}.

Many sources of non-Markovianity, such as strong system-reservoir coupling, being in contact with a structured bath, low environmental temperature, initial system-reservoir correlations, being driven by
classical noise~\cite{benedetti13,benedetti14,rossi16,lofranco14} and being in contact with
multiple reservoirs~\cite{man14}, have been found. Kutvonen et. al.~\cite{kutvonen15} suggested that the non-Markovianity could be accounted for by assuming the bath as combination of a part in thermal equilibrium and a part that is in non-equilibrium which does not change while the transitions in the system take place. Mixing of random unitary dynamics as well as being driven by classical noise have been shown to lead to
strong non-Markovian effects in both theoretical and experimental studies~\cite{benedetti13,benedetti14,rossi16,cialdi17,megier2017,breuer18}. There has been a discussion of whether non-Markovianity arising from a mixing of random unitary dynamics could also be considered as manifestation of backflow of information from the environment to the system~\cite{megier2017}
which was settled in positive~\cite{breuer18}. One of the manifestations of the memory effects of the dynamics has been seen in the entanglement revivals in which entanglement of the system is stored and back-transferred to the qubits by the quantum environment~\cite{bellomo07,bellomo08}. Because of the lack of back-action and the inability to store and share quantum correlations for the classical fields, the emergence of those revivals in quantum systems driven by classical random forces has been difficult to explain~\cite{lofranco14}. The effect of various noise types on the non-Markovianity of dynamics of qubit(s) have been studied by a number of groups in various settings~\cite{benedetti13,benedetti14,rossi16}. Ref.~\cite{benedetti14} investigated the effect of longitudinal telegraph noise which causes pure dephasing and $1/f^{\alpha}$ noise and have found that there exist a close connection between the non-Markovianity and the auto-correlation time of the noise. Rossi and Paris~\cite{rossi16} have shown that a transverse telegraph noise would lead to non-Markovian dynamics. Recently, Abel and Marquardt~\cite{Abel2008} investigated the dynamics of a charge qubit coupled to quantum telegraph noise and evaluated the time evolution of the coherence numerically. It is found that in the strong-coupling regime beyond a certain threshold the decay behavior of the coherence converts into an oscillation in time in contrary to the influence of any Gaussian noise source. Rossi and Paris have studied the dynamics of single- and two qubit system interacting with either Gaussian noise or random telegraph noise and analyzed the effect of these noises on the behavior of quantum correlations and found that both fast telegraph and Gaussian noises cause to decay incoherently for the quantum correlation, while the slow dichotomous noise allows it to oscillate intensively as sudden death and rebirth tendency, but the correlation is vanished rapidly under the effect of slow Gaussian environment~\cite{Rossi2016}. Man et al have shown that while the dynamics of a qubit coupled to a single bath can be Markovian or non-Markovian depending on the coupling strength, its dynamics is always non-Markovian if it is coupled to $N>N_c$ baths where $N_c$ depends on the bath parameters~\cite{man14}. Relation between the Markovianity of the dynamics and backflow of various physical properties, such as energy~\cite{liu17}, heat~\cite{schmidt16}, from the environment to the system were also considered.

In the present work, we investigate the non-Markovianity of the  dynamics of a two-state system (TSS) which is in contact with a thermal bath and its transition energy is driven by a classical telegraph noise. Such multi-environment couplings might arise in several contexts, for example, in long-range electron transfer processes in biological systems the solvent environment could be considered as collection of harmonic oscillators while the low frequency and/or large-amplitude motion of molecular environment (which can not be treated
in harmonic approximation) could be described as a two-state Markovian  noise~\cite{Gehlen1994,Petrov1994,Goychuk1995,Combined1997, May2003, Albinsson2008}. Coupling to both classical and quantum environments has been used to model the stochastic disturbances in the energy gap~\cite{Petrov1994,Goychuk1995,Combined1997}, electronic coupling~\cite{Tang1993,BridgeAssist1995,Dissipative1995,Control1997,Jan2000} and both the gap and the coupling~\cite{Kinetic1995} to investigate the effect of such motions on the transfer rate in spin-boson model. Stochastic driving of dissipative systems has been shown to violate detailed-balance condition which indicates  non-equilibrium dynamics~\cite{goychuk05}. So, the problem investigated in the
present report can also be considered as an example of a bath composed of a part which is in thermal equilibrium and a non-equilibrium part as Ref.~\cite{kutvonen15}. There has been discussion on the relation between the time-locality of the master equation used to describe the system dynamics and the non-Markovianity of the dynamics. Mazzola et al~\cite{mazzola10} have investigated the question of whether memory kernel master equations
always describe non Markovian dynamics as characterized by reverse information flow by calculating the BLP measure for a phenomenological and Shabani-Lidar post-Markovian master equations and have shown that $\mathcal{N}=0$ for both dynamics. We have, also, tested the  time-locality of the master equation
dependence of the non-Markovianity of the system dynamics in the present study and found that both the memory kernel
and the time convolutionless master equations produce similar non-Markovianity features at the considered
system parameters limits.

Outline of the present paper is as follows: In Section~\ref{model}, we present the model, the memory-kernel and time-convolutionless master equations, carry out the noise averaging of the dynamical equations and briefly describe the BLP non-Markovianity measure. Calculations on the non-Markovianity of the model with and without external noise as function of system and noise parameters are presented and discussed in Section~\ref{results}. Section~\ref{conc} concludes the paper with summary of the main findings.

\section{\label{model} Model}
We consider a dichotomously driven two state system (TSS) in contact with a thermal bath. The total Hamiltonian of the closed system formed by the TSS and its environment can be written as:
\begin{equation}
H=H_{S}(t)+H_{B}+H_{I}, \label{eq:HamEffectNoise}
\end{equation}
with
\begin{eqnarray}
H_{S}(t)&=&\frac{\epsilon\left(t\right)}{2}\sigma_{z}+\frac{V}{2}\sigma_{x}\label{eq:Hs},\\
H_{B}&=&\sum_{\lambda}\omega_{\lambda}b^{\dagger}_{\lambda}\,b_{\lambda}\label{eq:Hb},\\ H_{I}&=&\sum_{\lambda}g_{\lambda}\left(b^{\dagger}_{\lambda}+b_{\lambda}\right)\sigma_{z},\label{eq:int}
\end{eqnarray}
where $H_{S}\left(t\right)$, $H_{B}$, and $H_{I}$ describe the two level system driven by the telegraph noise, the bath which is composed of independent harmonic oscillators with natural frequencies $\omega_{\lambda}$  and the interaction between the system and the environment, respectively. Here $\sigma_{i}$ with $i=x,y,z$ are the Pauli spin matrices, $V$ is the tunneling splitting between the two states of the TSS.
$\epsilon\left(t\right)=\epsilon_{0}+\Omega\,\alpha\left(t\right)$ describes the dichotomously driven transition energy of the TSS with $\epsilon_{0}$ as its static value while $\Omega$ is the amplitude of the external stochastic field. $\alpha\left(t\right)$ describes the dichotomous Markov process (DMP) with possible values $\pm 1$ and the average $\langle \alpha\left(t\right)\rangle=0$. DMP autocorrelation has an  exponential decay form, \emph{i.e.,} $\langle \alpha(t)\alpha(t^{'})\rangle=\exp\left(-\nu\left|t-t^{'}\right|\right)$ where $\nu$ is the jumping rate of the noise. $b^{\dagger}_{\lambda}$ and $b_{\lambda}$ are the creation and annihilation operators of the environmental oscillators, while $g_{\lambda}$ denotes the coupling constant between the TSS and the $\lambda$th harmonic oscillator in the bath with frequency $\omega_{\lambda}$. The initial state of the closed system is assumed to be in the product form $\rho(0)=\rho_{S}(0)\otimes \rho_{B}(0)$ with the bath in thermal equilibrium at inverse temperature $\beta$ which leads to $\rho_{B}=\exp{\left(-\beta H_{B}\right)}/\mathrm{Tr}\left[\exp{\left(-\beta H_{B}\right)}\right]$ .

The effect of interaction between the TSS and
the harmonic bath is characterized by the bath spectral density
$J(\omega)=\sum_{\lambda}g_{\lambda}^2\delta\left(\omega-\omega_{\lambda}\right)$ which is assumed to be of the structured form~\cite{Hopfield1986,garg1985}
\begin{equation}\label{eq:spectral}
J(\omega)=8\,\kappa^2\,\frac{\gamma \,\omega_0\,\omega}{\left(\omega^2-\omega_0^2\right)^2+4 \gamma^2\omega^2}
\end{equation}
\noindent in the present study. Here $\kappa$ is the strength of the coupling between the TSS and the oscillator, $\omega_{0}$ is the center frequency of the bath oscillator, $\gamma$ indicates the broadening of the levels of the oscillator due to its environment.

In case of strong coupling between the system and the bath, the polaron transformation can be used to either decrease the efficiency of interaction term or destroy the system-bath coupling constant by transforming the total Hamiltonian to the polaron frame. The generator of this transformation is the operator:
\begin{equation}
\label{eq:polaron}
U=\frac{1}{2}\mathcal{B}\sigma_{z},
\end{equation}
\noindent where $\mathcal{B}=\sum_{\lambda}\frac{g_{\lambda}}{\omega_{\lambda}}\left(b^{\dagger}_{\lambda}
-b_{\lambda}\right)$. The transformation leads to a shift in the position of bath oscillator based on the state of the TSS. Applying this transformation to the Hamiltonian in Eq.~\ref{eq:HamEffectNoise}, one obtains:
\begin{eqnarray}
\label{eq:transHam}
H^{'}_{tot}&=&e^{U}\,H\,e^{-U}=H'_{0}+H'_{B}+H'_I,\nonumber\\
&=&\frac{\epsilon\left(t\right)}{2}\sigma_{z}+\frac{V_r}{2}\,\sigma_{x}+\sum_{\lambda}\omega_{\lambda}
b^{\dagger}_{\lambda}\,b_{\lambda}+\frac{V}{2}\left(\sigma_{+}\,B_{-}+\sigma_{-}\,B_{+}\right),
\end{eqnarray}
\noindent where $B_{\pm}=\exp{\left(\pm i \mathcal{B}\right)}$ and $\sigma_{\pm}=\left(\sigma_{x}\pm i\,\sigma_{y}\right)/2$ are the TSS raising and lowering operators of the system. Under polaron transformation, the environmental Hamiltonian does not change, the tunneling term of the system Hamiltonian is rescaled as $V\,e^{-Q(\tau)}$ which becomes zero for spectral densities that have a power exponent less than 2~\cite{Nazir2016,Goychuk1995}.

In polaron frame, because of $J(\omega)$ in Eq.~(\ref{eq:spectral}), $V_r=0$. Under those conditions, the system-environment interaction can be described in the form:
\begin{equation}\label{eq:HForm}
H'_{I}=\sum_{i}A_{i}\otimes B_{i},
\end{equation}
\noindent where $A_{1,2}=\{\sigma_{+},\sigma_{-}\}$ presents operators of the system, while $B_{1,2}=\{B_{-},B_{+}\}$ defines bath operators. A memory kernel master equation in the interaction picture for the density operator of the TSS can be derived for Hamiltonian of Eq.~(\ref{eq:transHam}) by using projector operator technique in Nakajima-Zwanzig form as follows~\cite{weiss98}:
\begin{eqnarray}\label{eq:NZME}
 \dot{\rho}_{S}'(t)&=&-\int_{0}^{t}dt_{1}\,\mathrm{Tr}_{\mathrm{B}}
 {\left[\bar{H}'_{I}(t),\left[\bar{H}'_{I}(t_{1}),
  \rho_{S}(t_1)\otimes\rho_{B}\right]\right]},
\end{eqnarray}
where $\mathrm{Tr}_{\mathrm{B}}$ indicates partial trace over the bath degrees of freedom, $\bar{H}'_{I}(t)=U_{0}\,H'_{I}\,U_{0}^{\dagger}$ with $U_{0}=e^{-i\int_{0}^{t}\,d\tau\,H'_{0}(\tau)}$ and $H'_{0}(t)=\frac{\epsilon\left(t\right)}{2}\sigma_{z}+\sum_{\lambda}\omega_{\lambda}
b^{\dagger}_{\lambda}\,b_{\lambda}$ is the interaction Hamiltonian in the interaction picture and polaron frame. Time convolutionless (TCL) master equation can be obtained from Eq.~(\ref{eq:NZME}) by simply changing the argument of $\rho_S(t)$ in the integrand from $t_1$ to $t$. Some model dependent studies indicate that TCL might describe system dynamics better than NZ.

 Polaron frame NZ and TCL form master equations for the system density operator $\rho_S(t)$ in the Shr\"{o}dinger picture can be derived with the help of Eq.~(\ref{eq:NZME}) as follows:
\begin{eqnarray}
\dot{\rho}^{NZ}_S(t)=-i\left[H_{S},\rho_{S}^{NZ}(t)\right]&&-\int_{0}^{t}d\,t'\sum_{i,i'}\{\left(A_{i}\,U(t,t')\,A_{i'}\,\rho_{S}^{NZ}(t')\,U^{\dagger}(t,t')\right.\nonumber\\
&&\left.-U(t,t')\,A_{i'}\,\rho_{S}^{NZ}(t')\,U^{\dagger}(t,t')\,A_{i}\right)\langle B_{i}(t)\,B_{i'}(t')\rangle\nonumber\\
&&+\left(U(t,t')\,\rho_{S}^{NZ}(t')\,A_{i'}\,U^{\dagger}(t,t')\,A_{i}\right.\nonumber\\
&&\left.-A_{i}\,U(t,t')\,\rho_{S}^{NZ}(t')\,A_{i'}\,U^{\dagger}(t,t')\right)\langle B_{i'}(t')\,B_{i}(t)\rangle\}
\label{eq:nzu}
\end{eqnarray}
\noindent and
\begin{eqnarray}
\dot{\rho}^{TCL}_S(t)=-i\left[H_{S},\rho_{S}^{TCL}(t)\right]&&-\int_{0}^{t}d\,t'\sum_{i,i'}\{\left(A_{i}\,U(t,t')\,A_{i'}\,U^{\dagger}(t,t')\,\rho_{S}^{TCL}(t)\right.\nonumber\\
&&\left.-U(t,t')\,A_{i'}\,U^{\dagger}(t,t')\,\rho_{S}^{TCL}(t)\,A_{i}\right)\langle B_{i}(t)\,B_{i'}(t')\rangle\nonumber\\
&&+\left(\rho_{S}^{TCL}(t)\,U(t,t')\,A_{i'}\,U^{\dagger}(t,t')\,A_{i}\right.\nonumber\\
&&\left.-A_{i}\,\rho_{S}^{TCL}(t)\,U(t,t')\,A_{i'}\,U^{\dagger}(t,t')\right)\langle B_{i'}(t')\,B_{i}(t)\rangle\}
\label{eq:tclu},
\end{eqnarray}
\noindent where the superscript on $\rho_S(t)$ in the left-hand side indicates the form of the master equation. In both equations, the propagator of the coherent system dynamics $U(t,t')$ can be expressed as:
\begin{eqnarray}
U\left(t,t'\right)&=&\mathcal{T}\left[
\exp{\left(-\frac{i}{2}\int_{t'}^{t}d\tau\;\left[\epsilon_{0}+\epsilon(\tau)\right]\sigma_{z}\right)}\right]
\nonumber\\
&=&\mathcal{I}\cos{\left[F\left(t,t'\right)/2\right]}+i\;\sigma_z\sin{\left[F\left(t,t'\right)/2\right]}
\label{eq:propagator},
\end{eqnarray}
\noindent where $\mathcal{T}$ indicates time-ordering, $\mathcal{I}$ is the $2\times 2$ unit matrix and
\begin{eqnarray}
F(t,t')=\epsilon_0\left(t-t'\right)+\Omega\int_{t'}^{t}d\tau \;\alpha(\tau)
\end{eqnarray}
\noindent contains both the static gap and the integral of the noise. We should note that after this point, we will work Shr\"{o}dinger picture and will drop both superscripts and over bars from the operators. Although the starting equations for NZ and TCL projections are quite similar (Eq.~(\ref{eq:NZME})), the final dynamical equations display a number of differences which will be discussed below.

The dynamics of the TSS in the memory kernel formulation can be obtained from Eq.~(\ref{eq:nzu}) with the help of the propagator $U(t,t')$ defined in Eq.~\ref{eq:propagator} by expressing the system density matrix as $\rho_{S}(t)=(\mathcal{I}+P_{i}(t)\cdot\sigma)/2$ where $P_{i}(t)=\mathrm{Tr_{S}}\left[\sigma_{i}\,\rho_{\mathrm{S}}(t)\right]$ as follows:
\begin{eqnarray}
\frac{d}{dt}P_{x}^{NZ}(t)&=&\left[\epsilon_{0}+\Omega\,\alpha(t)\right]P_{y}^{NZ}(t)-2\,V^{2}\int_{0}^{t}dt'\,e^{-Q_2(t-t')}\cos{\left[Q_{1}(t-t')\right]}\,P_{x}^{NZ}(t')\label{eq:PxNZ},\\
\frac{d}{dt}P_{y}^{NZ}(t)&=&-\left[\epsilon_{0}+\Omega\,\alpha(t)\right]P_{x}^{NZ}(t)-2\,V^{2}\int_{0}^{t}dt'\,e^{-Q_2(t-t')}\cos{\left[Q_{1}(t-t')\right]}\,P_{y}^{NZ}(t')\label{eq:PyNZ},\\
\frac{d}{dt}P_{z}^{NZ}(t)&=&-4\,V^{2}\int_{0}^{t}dt'\,e^{-Q_2(t-t')}\sin{\left[Q_{1}(t-t')\right]}\sin{\left[F(t,t')\right]}\nonumber\\
&&-4\,V^{2}\int_{0}^{t}dt'\,e^{-Q_2(t-t')}\cos{\left[Q_{1}(t-t')\right]}\cos{\left[F(t,t')\right]}\,P_{z}^{NZ}(t')\label{eq:PzNZ}.
\end{eqnarray}
Similarly, time convolutionless equations can be deduced from Eq.~(\ref{eq:tclu}) as:
\begin{eqnarray}
\frac{d}{dt}P_{x}^{TCL}(t)&=&\left[\epsilon_{0}+\Omega\,\alpha(t)\right]P_{y}^{TCL}(t)-2\,V^{2}\int_{0}^{t}dt'\,e^{-Q_2(t-t')}\cos{\left[Q_{1}(t-t')\right]}\,\cos{\left[F(t,t')\right]}\,P_{x}^{TCL}(t)\nonumber\\
&&+2\,V^{2}\int_{0}^{t}dt'\,e^{-Q_2(t-t')}\cos{\left[Q_{1}(t-t')\right]}\,\sin{\left[F(t,t')\right]}\,P_{y}^{TCL}(t)\label{eq:PxTCL},\\
\frac{d}{dt}P_{y}^{TCL}(t)&=&-\left[\epsilon_{0}+\Omega\,\alpha(t)\right]P_{x}^{TCL}(t)-2\,V^{2}\int_{0}^{t}dt'\,e^{-Q_2(t-t')}\cos{\left(Q_{1}(t-t')\right)}\,\sin{(F(t,t'))}\,P_{x}^{TCL}(t)\nonumber\\
&&-2\,V^{2}\int_{0}^{t}dt'\,e^{-Q_2(t-t')}\cos{\left(Q_{1}(t-t')\right)}\,\cos{(F(t,t'))}\,P_{y}^{TCL}(t)\label{eq:PyTCL},\\
\frac{d}{dt}P_{z}^{TCL}(t)&=&-4\,V^{2}\int_{0}^{t}dt'\,e^{-Q_2(t-t')}\sin{\left(Q_{1}(t-t')\right)}\sin{\left(F(t,t')\right)}\nonumber\\
&&-4\,V^{2}\int_{0}^{t}dt'\,e^{-Q_2(t-t')}\cos{\left(Q_{1}(t-t')\right)}\cos{\left(F(t,t')\right)}\,P_{z}^{TCL}(t)\label{eq:PzTCL},
\end{eqnarray}
\noindent where $Q_{1}(t)$ and $Q_{2}(t)$ are the imaginary and the real parts of the bath correlation function, respectively  and
are defined in terms of the bath spectral function as:
\begin{eqnarray}
Q_{1}(t)&=&\frac{1}{2\,\pi}\int_{0}^{\infty}d\omega\frac{J(\omega)}{\omega^{2}}\sin(\omega t)\label{eq:1q1q2},\\
Q_{2}(t)&=&\frac{1}{2\,\pi}\int_{0}^{\infty}d\omega\frac{J(\omega)}{\omega^{2}}\coth\left(\frac{\beta\,\omega}{2}\right)
\left(1-\cos(\omega t)\right) \label{eq:2q1q2}
\end{eqnarray}
\noindent which enter into dynamical equations \ref{eq:nzu} and \ref{eq:tclu} via average of bath operators $B_{\pm}$ as:
\begin{eqnarray}
\langle B_{\pm}(0)B_{\mp}(t)\rangle &=& e^{-Q_{2}(t)-i\,Q_{1}(t)},\\
\langle B_{\pm}(t)B_{\mp}(0)\rangle &=& e^{-Q_{2}(t)+i\,Q_{1}(t)}.
\end{eqnarray}
Although the memory kernel  and time-local equations are obtained from quite similar starting equations (\ref{eq:nzu} and \ref{eq:tclu}, respectively), an inspection of the resulting equations (\ref{eq:PxNZ}-\ref{eq:PzNZ}) and (\ref{eq:PxTCL}-\ref{eq:PzTCL}) indicates that populations and the coherences are independent of each other in both formulations. While the time-rate of change for the population are similar in form for both formulations (Eqs.~\ref{eq:PzNZ} and \ref{eq:PzTCL}) with the exception of the time-argument of $P_z(t)$ on the right-hand side, dynamical equations for the coherences in NZ and TCL formulations display a number of important differences. For instance, in NZ master equation for the coherences, the memory kernel depends on the environmental correlation functions only (Eqs.~\ref{eq:PxNZ} and \ref{eq:PyNZ}) while in TCL formulation the corresponding time-dependent coefficients include noise-effects as well as the TSS bias $\epsilon_0$ ($F(t,t')$ in Eqs.~\ref{eq:PxTCL} and \ref{eq:PyTCL}).

\subsection{Stochastic Averaging}

The dynamical equations \ref{eq:PxNZ}-\ref{eq:PzNZ} (NZ) and \ref{eq:PxTCL}-\ref{eq:PzTCL} (TCL) include stochastic terms $\alpha(t)$ and integral of $\alpha(t)$ and should be averaged over the realizations of the noise process which can be accomplished by either ensemble averaging, i.e. solving those equations for a large number of noise realizations and averaging the results or by averaging the set of coupled differential equations over the noise probability density. We will use the latter approach and follow the method of Ref.~\cite{Goychuk1995} which is based on Bourret-Frisch~\cite{Bourret73} and Shapiro-Loginov~\cite{Shapiro78} theorems and will make use of the results in Ref.~\cite{Luca2017}. Let $\langle P_{i}(t)\rangle$ be the noise averaged $P_i(t)$.
The dichotomous nature of the stochastic field $\alpha(t)$ makes it possible to carry out the averaging in exact form, but the number of coupled differential equations is doubled in the process; for each dynamical variable $\langle P_{i}(t)\rangle$ one needs to also find the evolution of $\langle\alpha_{i}(t)\rangle=\langle\alpha(t)\,P_{i}(t)\rangle$. We will present the results of the averaged NZ equations first:

\begin{eqnarray}
\frac{d}{dt}\langle P_{x}(t)\rangle&=&\epsilon_{0}\,\langle P_{y}(t)\rangle+\Omega\,\langle\alpha_{y}(t)\rangle-2\,V^{2}\,\int_{0}^{t}\,d\,t'\,e^{-Q_{2}(t-t')}\,\cos{\left(Q_1\left(t-t'\right)\right)}\langle P_x(t')\rangle,\label{eq:AvPxNZ}\\
\frac{d}{dt}\langle P_{y}(t)\rangle&=&-\epsilon_{0}\,\langle P_{x}(t)\rangle-\Omega\,\langle\alpha_{x}(t)\rangle-2\,V^{2}\,\int_{0}^{t}\,d\,t'\,e^{-Q_{2}(t-t')}\,\cos{\left(Q_1\left(t-t'\right)\right)}\langle P_y(t')\rangle,\label{eq:AvPyNZ}\\
\frac{d}{dt}\langle P_{z}(t)\rangle&=&-\int_{0}^{t}\,dt'\,K_1(t-t')-\int_{0}^{t}\,dt'\,K_2(t-t')\,\langle P_z(t')\rangle+\int_{0}^{t}\,dt'\,K_3(t-t')\,\langle\alpha_{z}(t')\rangle\label{eq:AvPzNZ},
\end{eqnarray}
\begin{eqnarray}
\frac{d}{dt}\langle \alpha_{x}(t)\rangle&=&-\nu\,\langle\alpha_{x}(t)\rangle+\epsilon_{0}\,\langle \alpha_{y}(t)\rangle+\Omega\,\langle P_{y}(t)\rangle\nonumber\\
&&-2\int_{0}^{t}\,d\,t'\,e^{-Q_{2}(t-t')}\,\cos{\left(Q_1\left(t-t'\right)\right)}\,e^{-\nu|t-t'|}\langle \alpha_x(t')\rangle,\label{eq:AvaPxNZ}\\
\frac{d}{dt}\langle \alpha_{y}(t)\rangle&=&-\nu\,\langle\alpha_{y}(t)\rangle-\epsilon_{0}\,\langle \alpha_{x}(t)\rangle-\Omega\,\langle P_{x}(t)\rangle\nonumber\\
&&-2\int_{0}^{t}\,d\,t'\,e^{-Q_{2}(t-t')}\,\cos{\left(Q_1\left(t-t'\right)\right)}\,e^{-\nu|t-t'|}\langle \alpha_y(t')\rangle,\label{eq:AvaPyNZ}\\
\frac{d}{dt}\langle\alpha_{z}(t)\rangle&=&-\nu\,\langle\alpha_z(t)\rangle+\int_{0}^{t}\,dt'\,K_3(t-t')\langle P_z(t')\rangle-\int_{0}^{t}\,dt'\,K_{4}(t-t')\nonumber\\
&&-\int_{0}^{t}\,dt'\,K_5(t-t')\langle\alpha_{z}(t')\rangle\label{eq:AvaPzNZ}.
\end{eqnarray}
\noindent where $K_{i}(t)$ are defined as
\begin{eqnarray*}
K_1(t)&=&4\,V^{2}\,e^{-Q_{2}(t)}\,\sin{\left(Q_{1}(t)\right)}\,S_{0}(t)\,\sin{\left(\epsilon_0(t)\right)},\\
K_2(t)&=&4\,V^{2}\,e^{-Q_{2}(t)}\,\cos{\left(Q_{1}(t)\right)}\,S_{0}(t)\,\cos{\left(\epsilon_0(t)\right)},\\
K_3(t)&=&4\,i\,V^{2}\,e^{-Q_{2}(t)}\,\cos{\left(Q_{1}(t)\right)}\,S_{1}(t)\,\sin{\left(\epsilon_0(t)\right)},\\
K_4(t)&=&4\,i\,V^{2}\,e^{-Q_{2}(t)}\,\sin{\left(Q_{1}(t)\right)}\,S_{1}(t)\,\cos{\left(\epsilon_0(t)\right)},\\
K_5(t)&=&4\,V^{2}\,e^{-Q_{2}(t)}\,\cos{\left(Q_{1}(t)\right)}\,S_{2}(t)\,\cos{\left(\epsilon_0(t)\right)},
\end{eqnarray*}
$S(t,t')=\exp\left[-i\,\Omega\int_{t'}^{t}d\tau\,\alpha(\tau)\right]$ is the time evolution operator of the Kubo oscillator and satisfies the stochastic evolution equation. In equations~(\ref{eq:AvPzNZ})-(\ref{eq:AvaPzNZ}), $S_{0}(t)$, $S_{1}(t)$, $S_{2}(t)$ are noise propagators of the dichotomous noise that can be evaluated by using $S(t,t')$ \cite{Goychuk1995} and are defined as:
\begin{eqnarray}
S_{0}(t)&=&\frac{1}{2\,\eta}\left(\nu_{+}\,e^{-t\,\nu_{-}/2}-\nu_{-}\,e^{-t\,\nu_{+}/2}\right),\label{eq:s0}\\
S_{1}(t)&=&i\frac{\Omega}{\nu}\left(e^{-t\,\nu_{+}/2}-e^{-t\,\nu_{-}/2}\right),\\
S_{2}(t)&=&\frac{1}{2\,\eta}\left(\nu_{+}\,e^{-t\,\nu_{+}/2}-\nu_{-}\,e^{-t\,\nu_{-}/2}\right),
\end{eqnarray}
\noindent where $\eta=\sqrt{\nu^{2}-4\,\Omega^{2}}$, $\nu_{+}=\nu+\eta$, and $\nu_{-}=\nu-\eta$.

Similarly, the time local equations (\ref{eq:PxTCL}-\ref{eq:PzTCL}) can be averaged over the dichotomous noise $\alpha(t)$ to obtain:
\begin{eqnarray}
\frac{d}{dt}\langle P_{x}(t)\rangle&=&-\frac{1}{2}\,\Gamma_{2}(t)\langle P_{x}(t)\rangle+\frac{1}{2}\,\Gamma_{3}(t)\langle \alpha_{x}(t)\rangle+(\epsilon_0+\Gamma_{5}(t))\langle P_y(t)\rangle\nonumber\\
&&+(\Omega+\Gamma_{6}(t))\,\langle \alpha_y(t)\rangle,\label{eq:AvPxTCL}\\
\frac{d}{dt}\langle P_{y}(t)\rangle&=&-\frac{1}{2}\,\Gamma_{2}(t)\langle P_{y}(t)\rangle+\frac{1}{2}\,\Gamma_{3}(t)\langle \alpha_{y}(t)\rangle-(\epsilon_0+\Gamma_{5}(t))\langle P_x(t)\rangle\nonumber\\
&&-(\Omega+\Gamma_{6}(t))\,\langle \alpha_x(t)\rangle,\label{eq:AvPyTCL}\\
\frac{d}{dt}\langle P_{z}(t)\rangle&=&-\Gamma_{1}(t)-\Gamma_{2}(t)\,\langle P_z(t)\rangle+\Gamma_{3}(t)\,\langle \alpha_{z}(t)\rangle\label{eq:AvPzTCL},
\end{eqnarray}
\begin{eqnarray}
\frac{d}{dt}\langle \alpha_{x}(t)\rangle&=&-\left[\nu+\frac{1}{2}\,\Gamma_{2}(t)\right]\langle \alpha_{x}(t)\rangle+\frac{1}{2}\,\Gamma_{3}(t)\langle P_{x}(t)\rangle+(\epsilon_0+\Gamma_{5}(t))\langle \alpha_y(t)\rangle\nonumber\\
&&+(\Omega+\Gamma_{6}(t))\,\langle P_y(t)\rangle,\label{eq:AvaPxTCL}\\
\frac{d}{dt}\langle \alpha_{y}(t)\rangle&=&-\left[\nu+\frac{1}{2}\,\Gamma_{2}(t)\right]\langle \alpha_{y}(t)\rangle+\frac{1}{2}\,\Gamma_{3}(t)\langle P_{y}(t)\rangle-(\epsilon_0+\Gamma_{5}(t))\langle \alpha_x(t)\rangle\nonumber\\
&&-(\Omega+\Gamma_{6}(t))\,\langle P_x(t)\rangle\label{eq:AvaPyTCL},\\
\frac{d}{dt}\langle\alpha_{z}(t)\rangle&=&-\left[\nu+\Gamma_{2}(t)\right]\,\langle\alpha_z(t)\rangle+\Gamma_{3}(t)\langle P_{z}(t)\rangle+\Gamma_{4}(t)\label{eq:AvaPzTCL},
\end{eqnarray}
where
\begin{eqnarray*}
\Gamma_{i}(t)&=&\int_{0}^{t}\,dt'\,K_{i}(t'), \quad i=1,2,3,4,\\
\Gamma_{5}(t)&=&\int_{0}^{t}\,dt'\,2\,V^{2}\,e^{-Q_{2}(t')}\,\cos{\left(Q_{1}(t')\right)}\,S_{0}(t')\,\sin{\left(\epsilon_0(t')\right)},\\
\Gamma_{6}(t)&=&2\,i\,V^{2}\,e^{-Q_{2}(t')}\,\cos{\left(Q_{1}(t')\right)}\,S_{1}(t')\,\cos{\left(\epsilon_0(t')\right)}.
\end{eqnarray*}

\subsection{Non-Markovianity Measure}
We employ the widely used  trace-distance based measure developed by Breuer, Laine and Piilo~\cite{breuer09,laine10} (BLP)
to investigate the non-Markovianity of the dynamics produced by the telegraph noise averaged Nakajima-Zwanzig (Eqs.~\ref{eq:AvPxNZ}-\ref{eq:AvaPzNZ}) and
TCL (Eqs.~\ref{eq:AvPxTCL}-\ref{eq:AvaPzTCL} ) master equations. BLP measure is defined in terms of the information flow $\sigma(\rho_1,\rho_2)=\frac{d}{dt}D(\rho_1,\rho_2)$
where
\begin{equation}
 D\left(\rho_1,\rho_2\right)=\frac{1}{2}\mathrm{Tr}\left|\rho_1-\rho_2\right|
 \label{eq:distinguishability},
\end{equation}
 \noindent where $D(\rho_1,\rho_2)$ is the distinguishability between states $\rho_1$ and $\rho_2$. A monotonously decreasing $D(\rho_1,\rho_2)$ is considered to be a sign of unidirectional flow
of information from the system to its environment signifying Markovian dynamics while positive $\sigma(\rho_1,\rho_2)$
in any time interval is considered as indication of information back-flow from the environment to the system. BLP
measure is defined as an optimization problem:
\begin{equation}
 \mathcal{N}=\max_{\rho_1,\rho_2}\int_{\sigma>0}\;\sigma\left(\rho_1,\rho_2\right)\;dt
 \label{eq:blpMeasure}
\end{equation}
\noindent over all the possible initial states. Wissmann et. al.~\cite{wissmann12} have shown that for the
BLP measure the optimal initial states $\rho_1(0)$ and $\rho_2(0)$ are always orthogonal and lie on the boundary of
the state space. We have used $|\psi_{1,2}(0)\rangle=\frac{1}{\sqrt{2}}\left(|0\rangle\pm|1\rangle\right)$ as the
initial states in the present work.

\section{\label{results}Results}

\begin{figure}[!ht]
\centering
\includegraphics[width=16cm]{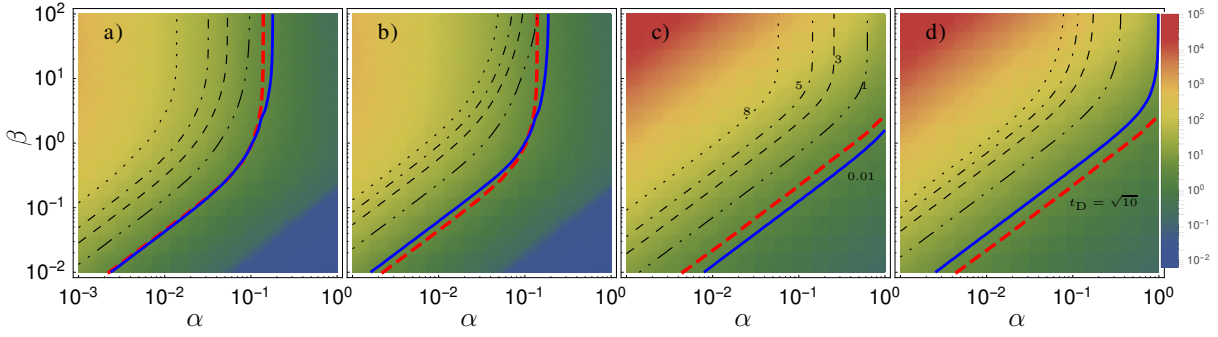}
\caption{Decay time of the bath correlation function and the non-Markovianity contours of the spin-boson model in
polaron frame as function of the dimensionless coupling constant $\alpha$ and the inverse temperature $\beta$ according
to the Nakajima-Zwanzig memory kernel master equation (a and c) and the
time convolutionless master equation (b and d) in under-(a and b $\gamma=0.1$) and over-damped
(c and d, $\gamma=100$) limits for the peak oscillator frequency $\omega_0=10$.
The graded color density plot and the legend bar
indicates the change of the bath correlation function decay time in logarithmic scales and the dashed red line shows the equality $\tau_{d}=\sqrt{10}$.}
\label{fig:nm-nonoise}
\end{figure}

We first discuss the non-Markovianity of the dynamics without the external noise. Figures~\ref{fig:nm-nonoise}a-d
 display the contour plots of BLP non-Markovianity measure on the density plot of the decay time of the environmental correlation function $\tau_d$ as function of the dimensionless interaction parameter $\alpha=4\,\kappa^{2}\,\gamma/\omega^{3}$ and the inverse temperature $\beta$ for the memory kernel Nakajima-Zwanzig  (Figs.~\ref{fig:nm-nonoise}a and \ref{fig:nm-nonoise}c) and the time convolutionless (Figs.~\ref{fig:nm-nonoise}b and \ref{fig:nm-nonoise}d) master equations at two different damping constants (under-damped limit $\gamma$=0.1 in Figs.~\ref{fig:nm-nonoise}a and \ref{fig:nm-nonoise}b and over-damped limit $\gamma$=100 in Figs.~\ref{fig:nm-nonoise}c and \ref{fig:nm-nonoise}d) for  $\epsilon_0=1$ and $\omega_0=10$.
We have defined the decay time $\tau_d$ as an estimation of the decay coefficient of the kernel function $G(t)=\exp{\left[-Q_2(t)\right]}$. Depending on the relative values of
$\omega_0$, $\gamma$, $\alpha$ and $\beta$, $G(t)$ might display  damped oscillations, resurgent damped oscillations or pure decaying behavior which is exponential or Gaussian in time. The displayed $\tau_d$ is obtained by fitting either $G(t)$ or its maxima to function $\exp{\left(-t/\tau_d\right)}$ or $\exp{\left(-t^2/\tau_d^2\right)}$ depending on whether $G(t)$ is monotonous in time or its time dependence display damped oscillations, respectively. As expected, the decay is the fastest when the system-environment coupling is large and the temperature is high ($\tau_d\approx 10^{-2}$). In this limit, the so-called short-time approximation~\cite{garg1985} can be used to express the environment correlation function as $G(t)=\exp{\left(-t^2/\tau_d^2\right)}$ with $\tau_d= \sqrt{\beta/E_r}$ where $E_r=\kappa^{2}/\omega_{0}$ is the reorganization energy of the system. $\tau_{d}$ is found to be the largest at the opposite limit of low temperature and weak system-environment coupling independent of the damping of the oscillator. As can be seen from a comparison of Figs.~\ref{fig:nm-nonoise}a and ~\ref{fig:nm-nonoise}c, $\tau_d$ is higher for the over-damped case compared to the under-damped limit at the same $\alpha$ and $\beta$ values. One should note that, sometimes, the inverse of the width of the spectral function $J(\omega)$ ($1/\gamma$) is considered as a measure of the correlation time of the environment.  Contrary to expectations, $\tau_{d}$ defined above seems to be directly proportional to $\gamma$.

The most important finding concerning the non-Markovianity of the dynamics of the TSS when there is no external noise is the close connection between the existence of non-Markovianity and the magnitude of $\tau_d$.
From figures ~\ref{fig:nm-nonoise}a-d it is obvious that $\mathcal{N}$ is non-zero when $\tau_d>\approx \sqrt{10}$ independent of the damping for both TCL and NZ master equations. In this regime,  the  characteristic time of the system dynamics is faster compared to the decay of the bath correlations which enables the information on the system to be retained in the environment and flow back into the system. Furthermore, the gross-features of the non-Markovianity of the NZ and TCL formulations seem to be similar for the parameters considered in the present work (the similarity depends on $\omega_0$ being large). A general observation from figures~\ref{fig:nm-nonoise}a-d is that the dynamics are non-Markovian at low temperatures almost independent of the other problem parameters and the type of master equation one uses to describe the system dynamics which is similar to the findings of Rivas who has shown that the spin-boson model approaches "eternal" non-Markovian regime as the temperature of the environment approaches zero~\cite{rivas17}. These findings are in  disagreement (agreement) with those reported by Chen et. al.~\cite{chen15} (Liu et.al.~\cite{liu15}) who reported that the increasing (decreasing) temperature and interaction strength increases the non-Markovianity for a spin-boson model in the context of photosynthetic systems.

\begin{figure}[!ht]
\centering
\includegraphics[width=16cm]{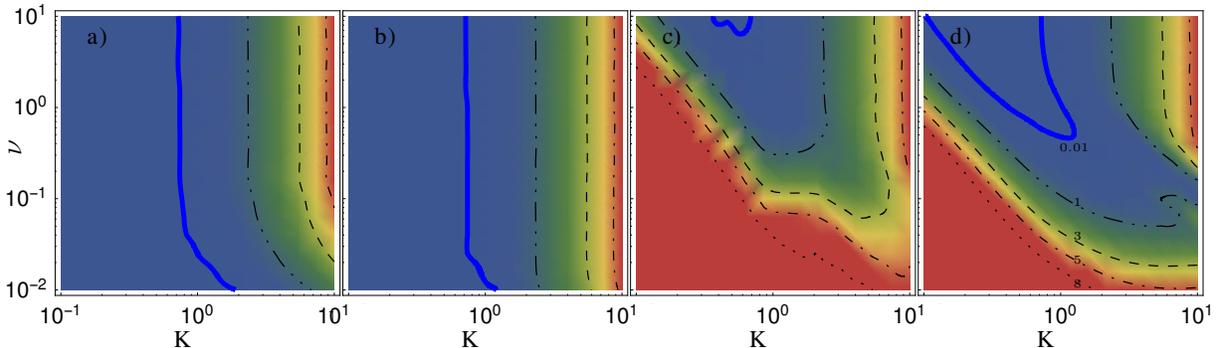}
\caption{Effect of the dichotomous noise on the non-Markovianity of the spin-boson model in
polaron frame as function of noise color $\mathrm{K}=\Omega/\nu$ and the noise frequency $\nu$ in the
Nakajima-Zwanzig memory kernel (a and c) and the time convolutionless master equation (b and d) approaches for weak coupling ($\alpha=0.0035$) at low temperature ($\beta=20$) (a and b) and strong-coupling $\alpha=0.35$ at high temperature $\beta=0.05$ (c and d, $\gamma=100$) limits for the peak oscillator frequency $\omega_0=10$.}
\label{fig:w003}
\end{figure}

We next investigate the effect of the telegraph noise on the non-Markovianity of the TSS dynamics in various points of the parameters space explored in figure~\ref{fig:nm-nonoise}. In figures~\ref{fig:w003}a-d, we present the non-Markovianity as function of the noise color $K=\Omega/\nu$ and the noise frequency in both NZ (\ref{fig:w003}a and \ref{fig:w003}c) and TCL (\ref{fig:w003}b and \ref{fig:w003}d) formulations. First, we consider the high temperature $(\beta=0.05)$, under-damped $(\gamma=0.1)$ and strong coupling $(\alpha=0.35)$ case for $\omega_0=10$ which is Markovian when there is no external noise. As can be seen from Figs.~\ref{fig:w003}a and \ref{fig:w003}b, the external noise creates non-Markovianity when the noise is slow ($K>1/2$, i.e. the noise propagator $S_0$ of Eq.~\ref{eq:s0} is oscillatory) in both NZ and TCL approaches and the magnitude of $\mathcal{N}$ increases with increasing $K$ with a weak-dependence on the noise frequency. A similar findings has been reported for the dichotomously driven qubit dynamics~\cite{benedetti14}. Figures~\ref{fig:w003}a and \ref{fig:w003}b, also, show that the fast jumping noise ($K\ll1$) does not create any non-Markovianity. The effect of noise on $\mathcal{N}$ is quite different when the dynamics is already non-Markovian in the absence of the external noise as can be seen from Figs.~\ref{fig:w003}c and ~\ref{fig:w003}d which display $\mathcal{N}$ for low temperature $(\beta=20)$, weak coupling $(\alpha=0.0035)$, over-damped $(\gamma=100)$. In those figures, $\mathcal{N}$ is always less than its non-noisy value and the effect of noise depends both on its frequency and the color. At high $\nu$ and intermediate $K$ values $\mathcal{N}$ approaches zero, while for strongly colored noise, $\mathcal{N}$ increases with increasing $K$ similar to the first case considered above which can be understood as the external noise effect dominating the thermal fluctuations of the environment.

\begin{figure}[!ht]
\centering
\includegraphics[width=14cm]{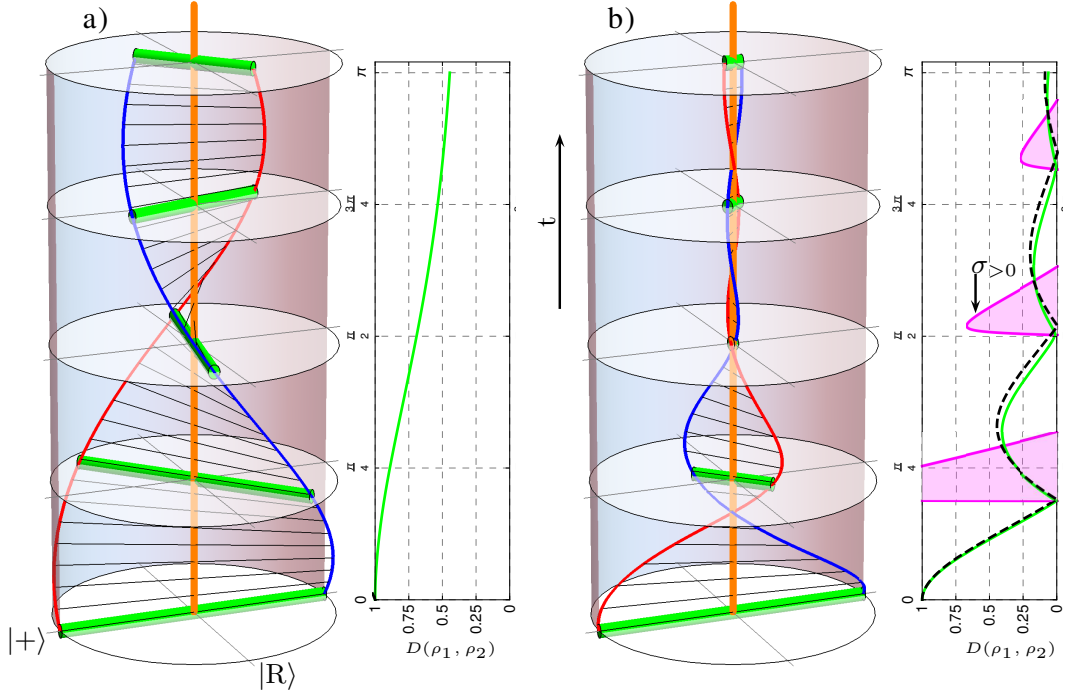}
\caption{Dynamics of $\rho^{(1)}$ and $\rho^{(2)}$ at $\epsilon_0=1$, $\omega_0=10$,
$\gamma=1$, $\alpha=0.035$, $\beta=0.1$, without external noise (a) and with external telegraph noise with
parameters $\nu=\pi/2$ and $\Omega=\pi$ (b). The line plots show the change in distinguishability with time, dashed line in b) is $D(\rho_1,\rho_2)$ calculated from Eq.\ref{eq:distin} and the shaded area displays the region where $dD(\rho_1,\rho_2)/dt>0$.}
\label{fig:w002}
\end{figure}

To further investigate the effect of noise on $\mathcal{N}$, we display the dynamics of coherences along with the distinguishability for parameters that lead to Markovian dynamics when there is no external noise and non-Markovian dynamics under the noise in Figs~\ref{fig:w002}a and \ref{fig:w002}b, respectively. Only TCL results are shown in the figure, the NZ ones are very similar. For the considered parameters of the problem and the initial states used in the calculation of $\mathcal{N}$ ($P_x(0)=\pm 1$), the change in $P_z$ is negligible
and the dynamics of the TSS can be depicted in cylindrical coordinates with $z-$axis representing time. Figures~\ref{fig:w002}a-b  shows the dynamics of the system with two chosen initial states in NZ and TCL approaches under no-noise and with external noise-driving with parameters $\nu=\pi/2$ and $\Omega=\pi$ which corresponds to $K=\Omega/\nu=2$ slow noise limit. From the figure~\ref{fig:w002}a  it is obvious that for the given parameters, the dynamics when there is no external noise are Markovian because the distinguishability, which is the distance between the two solutions shown as the lines connecting those solutions in the figure, decreases monotonously as time increases. On the other hand, external noise not only increases the decay of $P_x(t)$ and $P_y(t)$, as expected, but it also destroys the monotonicity of the distinguishability between $\rho_1(t)$ and $\rho_2(t)$ which leads to non-Markovianity. To delineate the source of such noise induced non-Markovianity, we will examine NZ and TCL master equations for the coherences Eqs.~(\ref{eq:AvPxNZ}-\ref{eq:AvPyNZ}, \ref{eq:AvaPxNZ}-\ref{eq:AvaPyNZ}) and Eqs.~(\ref{eq:AvPxTCL}-\ref{eq:AvPyTCL}, \ref{eq:AvaPxTCL}-\ref{eq:AvaPyTCL}), respectively.

For the strong coupling and the high temperature limit, both NZ and TCL equations for $\langle P_x(t)\rangle$ and
$\langle P_y(t)\rangle$ (Eqs.~\ref{eq:AvPxTCL} and \ref{eq:AvPyTCL}) and their noise correlators (Eqs.~\ref{eq:AvaPxTCL} and \ref{eq:AvaPyTCL}) can be approximated as:
\begin{eqnarray}
 \dot{P_x}(t)&=&\epsilon_0 P_{y}(t)+\Omega\, \alpha_{y}(t)  \nonumber \\
 \dot{P_y}(t)&=&-\epsilon_0 P_{x}(t)-\Omega\, \alpha_{x}(t) \nonumber \\
 \dot{\alpha_x}(t)&=&-\nu \alpha_x(t)+\epsilon_0 \alpha_{y}(t)+\Omega\, P_{y}(t)
 \nonumber \\
 \dot{\alpha_y}(t)&=&-\nu \alpha_y(t)-\epsilon_0 \alpha_{x}(t)-\Omega\, P_{x}(t)
 \label{eq:tclnz}
\end{eqnarray}
\noindent which can be solved analytically for the optimal initial states to obtain
the distinguishability as:
\begin{eqnarray}
 D(\rho_1,\rho_2)=\left|
 \frac{1}{2\eta}\left(\nu_{+}\mathrm{e}^{-\nu_{-}t/2}-\nu_{-}\mathrm{e}^{-\nu_{+}t/2}\right)\right|
\label{eq:distin}
\end{eqnarray}
\noindent which is same as the absolute value of the noise propagator $S_{0}$ (Eq.~\ref{eq:s0}). Note that although the static bias of the TSS $\epsilon_0$ enters into dynamical
equations \ref{eq:tclnz}, the distinguishability is independent of it. Using the definition of information
flow $\sigma(\rho_1,\rho_2)=\frac{dD(\rho_1,\rho_2)}{dt}$ and the BLP non-Markovianity measure
Eq.~\ref{eq:blpMeasure}, $\mathcal{N}$ can be obtained analytically from Eq.~\ref{eq:distin} as
\begin{equation}
\mathcal{N}=2\left(1-\frac{\nu(\nu+2)}{4 \Omega^2}\right)\frac{1}{\mathrm{e}^{\xi}-1}
\end{equation}
\noindent where $\xi=-2i\pi\nu/\eta$.
\noindent A similar expression was obtained for the non-Markovianity of the dynamics of a TSS under the influence of telegraph noise only by Ref.~\cite{benedetti14}.

It is interesting to note that even when the dynamical equations for the coherences are in the explicitly Markovian form as in Eq.~\ref{eq:tclnz}, they might describe a non-Markovian dynamics as measured by the change in the trace-distance based distinguishability when the external noise is slow. A similar finding is reported in Ref.~\cite{clos12} which has questioned the notion of standard Markov approximation by showing that the
master equation with stationary rates which is often regarded
as Markovian description does not necessarily lead to
a Markovian dynamics in the sense of unidirectional information flow from the system to the environment for the spin-boson model.  As the effect of slow noise in this context might be considered as creating a superposition  of two possible solutions with $\epsilon=\epsilon_0\pm \Omega$, our findings can be related with random mixtures of Markovian dynamical maps creating non-Markovian dynamics discussed in Ref.~\cite{breuer18}. Noise induced non-Markovianity can also be attributed to the non-equilibrium effects due to the stochastic driving which was shown to violate detailed-balance condition and lead to non-equilibrium dynamics~\cite{goychuk05} . Kutvonen et. al.~\cite{kutvonen15} suggested that the non-Markovianity could be accounted for by assuming the bath as combination of a part in thermal equilibrium and a part that is in non-equilibrium which does not change while the transitions in the system take place. So, the dichotomous driving in the present problem can also be considered as an example of non-equilibrium generating source.

\section{\label{conc}Conclusion}
We have considered the effect of dichotomous noise on the dynamics of a two state system which is in contact with a thermal bath of harmonic oscillators in the strong system-bath coupling regime in memory kernel and time local master equation approaches. The noise was assumed to modulate the transition energy of the TSS. We have obtained and numerically solved noise-averaged dynamical equations for both NZ and TCL approaches to study non-Markovianity of the TSS dynamics as quantified by the distinguishability based BLP measure. In the absence of the external noise, we have found that the non-Markovianity is strongly correlated with the decay time of the environmental correlation function which increases with decreasing coupling between the system and the environmental oscillator and the temperature of the bath and increasing of the damping of the oscillator. External noise is found to effect the non-Markovianity in two different ways depending on the Markovianity of the dynamics in the absence of the noise. When the dynamics are already non-Markovian, low frequency and weak external noise causes slight decrease in non-Markovianity while high frequency, intermediate noise makes the dynamics Markovian. On the other hand, slow noise is found to induce non-Markovianity when the dynamics is Markovian in its absence. Both memory kernel and time local formulations of the master equation are found to
signal similar BLP non-Markovianity features for the noiseless and noisy conditions which indicates that the form of the master equation is not a factor determining the Markovianity properties of the dynamics. Furthermore, although the dynamical equations we have obtained for the strong coupling, high temperature limit under the external noise driving have time-independent coefficients, they lead to non-Markovian dynamics which can be considered as
an example of random mixing induced non-Markovianity.


\newpage

\begin{thebibliography}{56}%
\makeatletter
\providecommand \@ifxundefined [1]{%
 \@ifx{#1\undefined}
}%
\providecommand \@ifnum [1]{%
 \ifnum #1\expandafter \@firstoftwo
 \else \expandafter \@secondoftwo
 \fi
}%
\providecommand \@ifx [1]{%
 \ifx #1\expandafter \@firstoftwo
 \else \expandafter \@secondoftwo
 \fi
}%
\providecommand \natexlab [1]{#1}%
\providecommand \enquote  [1]{``#1''}%
\providecommand \bibnamefont  [1]{#1}%
\providecommand \bibfnamefont [1]{#1}%
\providecommand \citenamefont [1]{#1}%
\providecommand \href@noop [0]{\@secondoftwo}%
\providecommand \href [0]{\begingroup \@sanitize@url \@href}%
\providecommand \@href[1]{\@@startlink{#1}\@@href}%
\providecommand \@@href[1]{\endgroup#1\@@endlink}%
\providecommand \@sanitize@url [0]{\catcode `\\12\catcode `\$12\catcode
  `\&12\catcode `\#12\catcode `\^12\catcode `\_12\catcode `\%12\relax}%
\providecommand \@@startlink[1]{}%
\providecommand \@@endlink[0]{}%
\providecommand \url  [0]{\begingroup\@sanitize@url \@url }%
\providecommand \@url [1]{\endgroup\@href {#1}{\urlprefix }}%
\providecommand \urlprefix  [0]{URL }%
\providecommand \Eprint [0]{\href }%
\providecommand \doibase [0]{http://dx.doi.org/}%
\providecommand \selectlanguage [0]{\@gobble}%
\providecommand \bibinfo  [0]{\@secondoftwo}%
\providecommand \bibfield  [0]{\@secondoftwo}%
\providecommand \translation [1]{[#1]}%
\providecommand \BibitemOpen [0]{}%
\providecommand \bibitemStop [0]{}%
\providecommand \bibitemNoStop [0]{.\EOS\space}%
\providecommand \EOS [0]{\spacefactor3000\relax}%
\providecommand \BibitemShut  [1]{\csname bibitem#1\endcsname}%
\let\auto@bib@innerbib\@empty
\bibitem [{\citenamefont {Rivas}\ \emph {et~al.}(2014)\citenamefont {Rivas},
  \citenamefont {Huelga},\ and\ \citenamefont {Plenio}}]{rivas14}%
  \BibitemOpen
  \bibfield  {author} {\bibinfo {author} {\bibfnamefont {A.}~\bibnamefont
  {Rivas}}, \bibinfo {author} {\bibfnamefont {S.~F.}\ \bibnamefont {Huelga}}, \
  and\ \bibinfo {author} {\bibfnamefont {M.~B.}\ \bibnamefont {Plenio}},\
  }\href@noop {} {\bibfield  {journal} {\bibinfo  {journal} {Reports on
  Progress in Physics}\ }\textbf {\bibinfo {volume} {77}},\ \bibinfo {pages}
  {094001} (\bibinfo {year} {2014})}\BibitemShut {NoStop}%
\bibitem [{\citenamefont {Breuer}\ \emph {et~al.}(2016)\citenamefont {Breuer},
  \citenamefont {Laine}, \citenamefont {Piilo},\ and\ \citenamefont
  {Vacchini}}]{breuer16}%
  \BibitemOpen
  \bibfield  {author} {\bibinfo {author} {\bibfnamefont {H.-P.}\ \bibnamefont
  {Breuer}}, \bibinfo {author} {\bibfnamefont {E.-M.}\ \bibnamefont {Laine}},
  \bibinfo {author} {\bibfnamefont {J.}~\bibnamefont {Piilo}}, \ and\ \bibinfo
  {author} {\bibfnamefont {B.}~\bibnamefont {Vacchini}},\ }\href@noop {}
  {\bibfield  {journal} {\bibinfo  {journal} {Rev. Mod. Phys.}\ }\textbf
  {\bibinfo {volume} {88}},\ \bibinfo {pages} {021002} (\bibinfo {year}
  {2016})}\BibitemShut {NoStop}%
\bibitem [{\citenamefont {de~Vega}\ and\ \citenamefont
  {Alonso}(2017)}]{devega17}%
  \BibitemOpen
  \bibfield  {author} {\bibinfo {author} {\bibfnamefont {I.}~\bibnamefont
  {de~Vega}}\ and\ \bibinfo {author} {\bibfnamefont {D.}~\bibnamefont
  {Alonso}},\ }\href@noop {} {\bibfield  {journal} {\bibinfo  {journal} {Rev.
  Mod. Phys.}\ }\textbf {\bibinfo {volume} {89}},\ \bibinfo {pages} {015001}
  (\bibinfo {year} {2017})}\BibitemShut {NoStop}%
\bibitem [{\citenamefont {Breuer}\ \emph {et~al.}(2009)\citenamefont {Breuer},
  \citenamefont {Laine},\ and\ \citenamefont {Piilo}}]{breuer09}%
  \BibitemOpen
  \bibfield  {author} {\bibinfo {author} {\bibfnamefont {H.-P.}\ \bibnamefont
  {Breuer}}, \bibinfo {author} {\bibfnamefont {E.-M.}\ \bibnamefont {Laine}}, \
  and\ \bibinfo {author} {\bibfnamefont {J.}~\bibnamefont {Piilo}},\
  }\href@noop {} {\bibfield  {journal} {\bibinfo  {journal} {Phys. Rev. Lett.}\
  }\textbf {\bibinfo {volume} {103}},\ \bibinfo {pages} {210401} (\bibinfo
  {year} {2009})}\BibitemShut {NoStop}%
\bibitem [{\citenamefont {Laine}\ \emph {et~al.}(2010)\citenamefont {Laine},
  \citenamefont {Piilo},\ and\ \citenamefont {Breuer}}]{laine10}%
  \BibitemOpen
  \bibfield  {author} {\bibinfo {author} {\bibfnamefont {E.-M.}\ \bibnamefont
  {Laine}}, \bibinfo {author} {\bibfnamefont {J.}~\bibnamefont {Piilo}}, \ and\
  \bibinfo {author} {\bibfnamefont {H.-P.}\ \bibnamefont {Breuer}},\
  }\href@noop {} {\bibfield  {journal} {\bibinfo  {journal} {Phys. Rev. A}\
  }\textbf {\bibinfo {volume} {81}},\ \bibinfo {pages} {062115} (\bibinfo
  {year} {2010})}\BibitemShut {NoStop}%
\bibitem [{\citenamefont {Rivas}\ \emph {et~al.}(2010)\citenamefont {Rivas},
  \citenamefont {Huelga},\ and\ \citenamefont {Plenio}}]{rivas10}%
  \BibitemOpen
  \bibfield  {author} {\bibinfo {author} {\bibfnamefont {A.}~\bibnamefont
  {Rivas}}, \bibinfo {author} {\bibfnamefont {S.~F.}\ \bibnamefont {Huelga}}, \
  and\ \bibinfo {author} {\bibfnamefont {M.~B.}\ \bibnamefont {Plenio}},\
  }\href@noop {} {\bibfield  {journal} {\bibinfo  {journal} {Phys. Rev. Lett.}\
  }\textbf {\bibinfo {volume} {105}},\ \bibinfo {pages} {050403} (\bibinfo
  {year} {2010})}\BibitemShut {NoStop}%
\bibitem [{\citenamefont {Benedetti}\ \emph {et~al.}(2013)\citenamefont
  {Benedetti}, \citenamefont {Buscemi}, \citenamefont {Bordone},\ and\
  \citenamefont {Paris}}]{benedetti13}%
  \BibitemOpen
  \bibfield  {author} {\bibinfo {author} {\bibfnamefont {C.}~\bibnamefont
  {Benedetti}}, \bibinfo {author} {\bibfnamefont {F.}~\bibnamefont {Buscemi}},
  \bibinfo {author} {\bibfnamefont {P.}~\bibnamefont {Bordone}}, \ and\
  \bibinfo {author} {\bibfnamefont {M.~G.~A.}\ \bibnamefont {Paris}},\
  }\href@noop {} {\bibfield  {journal} {\bibinfo  {journal} {Phys. Rev. A}\
  }\textbf {\bibinfo {volume} {87}},\ \bibinfo {pages} {052328} (\bibinfo
  {year} {2013})}\BibitemShut {NoStop}%
\bibitem [{\citenamefont {Benedetti}\ \emph {et~al.}(2014)\citenamefont
  {Benedetti}, \citenamefont {Paris},\ and\ \citenamefont
  {Maniscalco}}]{benedetti14}%
  \BibitemOpen
  \bibfield  {author} {\bibinfo {author} {\bibfnamefont {C.}~\bibnamefont
  {Benedetti}}, \bibinfo {author} {\bibfnamefont {M.~G.~A.}\ \bibnamefont
  {Paris}}, \ and\ \bibinfo {author} {\bibfnamefont {S.}~\bibnamefont
  {Maniscalco}},\ }\href@noop {} {\bibfield  {journal} {\bibinfo  {journal}
  {Phys. Rev. A}\ }\textbf {\bibinfo {volume} {89}},\ \bibinfo {pages} {012114}
  (\bibinfo {year} {2014})}\BibitemShut {NoStop}%
\bibitem [{\citenamefont {Rossi}\ and\ \citenamefont
  {Paris}(2016{\natexlab{a}})}]{rossi16}%
  \BibitemOpen
  \bibfield  {author} {\bibinfo {author} {\bibfnamefont {M.~A.~C.}\
  \bibnamefont {Rossi}}\ and\ \bibinfo {author} {\bibfnamefont {M.~G.~A.}\
  \bibnamefont {Paris}},\ }\href@noop {} {\bibfield  {journal} {\bibinfo
  {journal} {The Journal of Chemical Physics}\ }\textbf {\bibinfo {volume}
  {144}},\ \bibinfo {pages} {024113} (\bibinfo {year}
  {2016}{\natexlab{a}})}\BibitemShut {NoStop}%
\bibitem [{\citenamefont {Clos}\ and\ \citenamefont {Breuer}(2012)}]{clos12}%
  \BibitemOpen
  \bibfield  {author} {\bibinfo {author} {\bibfnamefont {G.}~\bibnamefont
  {Clos}}\ and\ \bibinfo {author} {\bibfnamefont {H.-P.}\ \bibnamefont
  {Breuer}},\ }\href@noop {} {\bibfield  {journal} {\bibinfo  {journal} {Phys.
  Rev. A}\ }\textbf {\bibinfo {volume} {86}},\ \bibinfo {pages} {012115}
  (\bibinfo {year} {2012})}\BibitemShut {NoStop}%
\bibitem [{\citenamefont {Haikka}\ \emph {et~al.}(2013)\citenamefont {Haikka},
  \citenamefont {Johnson},\ and\ \citenamefont {Maniscalco}}]{haikka13}%
  \BibitemOpen
  \bibfield  {author} {\bibinfo {author} {\bibfnamefont {P.}~\bibnamefont
  {Haikka}}, \bibinfo {author} {\bibfnamefont {T.~H.}\ \bibnamefont {Johnson}},
  \ and\ \bibinfo {author} {\bibfnamefont {S.}~\bibnamefont {Maniscalco}},\
  }\href@noop {} {\bibfield  {journal} {\bibinfo  {journal} {Phys. Rev. A}\
  }\textbf {\bibinfo {volume} {87}},\ \bibinfo {pages} {010103} (\bibinfo
  {year} {2013})}\BibitemShut {NoStop}%
\bibitem [{\citenamefont {Addis}\ \emph {et~al.}(2014)\citenamefont {Addis},
  \citenamefont {Bylicka}, \citenamefont {Chruscinski},\ and\ \citenamefont
  {Maniscalco}}]{addis14}%
  \BibitemOpen
  \bibfield  {author} {\bibinfo {author} {\bibfnamefont {C.}~\bibnamefont
  {Addis}}, \bibinfo {author} {\bibfnamefont {B.}~\bibnamefont {Bylicka}},
  \bibinfo {author} {\bibfnamefont {D.}~\bibnamefont {Chruscinski}}, \ and\
  \bibinfo {author} {\bibfnamefont {S.}~\bibnamefont {Maniscalco}},\
  }\href@noop {} {\bibfield  {journal} {\bibinfo  {journal} {Phys. Rev. A}\
  }\textbf {\bibinfo {volume} {90}},\ \bibinfo {pages} {052103} (\bibinfo
  {year} {2014})}\BibitemShut {NoStop}%
\bibitem [{\citenamefont {Lo~Franco}\ \emph {et~al.}(2014)\citenamefont
  {Lo~Franco}, \citenamefont {D'Arrigo}, \citenamefont {Falci}, \citenamefont
  {Compagno},\ and\ \citenamefont {Paladino}}]{lofranco14}%
  \BibitemOpen
  \bibfield  {author} {\bibinfo {author} {\bibfnamefont {R.}~\bibnamefont
  {Lo~Franco}}, \bibinfo {author} {\bibfnamefont {A.}~\bibnamefont {D'Arrigo}},
  \bibinfo {author} {\bibfnamefont {G.}~\bibnamefont {Falci}}, \bibinfo
  {author} {\bibfnamefont {G.}~\bibnamefont {Compagno}}, \ and\ \bibinfo
  {author} {\bibfnamefont {E.}~\bibnamefont {Paladino}},\ }\href@noop {}
  {\bibfield  {journal} {\bibinfo  {journal} {Phys. Rev. B}\ }\textbf {\bibinfo
  {volume} {90}},\ \bibinfo {pages} {054304} (\bibinfo {year}
  {2014})}\BibitemShut {NoStop}%
\bibitem [{\citenamefont {Schmidt}\ \emph {et~al.}(2016)\citenamefont
  {Schmidt}, \citenamefont {Maniscalco},\ and\ \citenamefont
  {Ala-Nissila}}]{schmidt16}%
  \BibitemOpen
  \bibfield  {author} {\bibinfo {author} {\bibfnamefont {R.}~\bibnamefont
  {Schmidt}}, \bibinfo {author} {\bibfnamefont {S.}~\bibnamefont {Maniscalco}},
  \ and\ \bibinfo {author} {\bibfnamefont {T.}~\bibnamefont {Ala-Nissila}},\
  }\href@noop {} {\bibfield  {journal} {\bibinfo  {journal} {Phys. Rev. A}\
  }\textbf {\bibinfo {volume} {94}},\ \bibinfo {pages} {010101} (\bibinfo
  {year} {2016})}\BibitemShut {NoStop}%
\bibitem [{\citenamefont {{Liu}}\ \emph {et~al.}(2017)\citenamefont {{Liu}},
  \citenamefont {{Xu}},\ and\ \citenamefont {{Wu}}}]{liu17}%
  \BibitemOpen
  \bibfield  {author} {\bibinfo {author} {\bibfnamefont {J.}~\bibnamefont
  {{Liu}}}, \bibinfo {author} {\bibfnamefont {H.}~\bibnamefont {{Xu}}}, \ and\
  \bibinfo {author} {\bibfnamefont {C.}~\bibnamefont {{Wu}}},\ }\href@noop {}
  {\bibfield  {journal} {\bibinfo  {journal} {ArXiv e-prints}\ } (\bibinfo
  {year} {2017})},\ \Eprint {http://arxiv.org/abs/1701.04570} {arXiv:1701.04570
  [quant-ph]} \BibitemShut {NoStop}%
\bibitem [{\citenamefont {Chen}\ \emph {et~al.}(2014)\citenamefont {Chen},
  \citenamefont {Lien}, \citenamefont {Hwang},\ and\ \citenamefont
  {Chen}}]{chen14}%
  \BibitemOpen
  \bibfield  {author} {\bibinfo {author} {\bibfnamefont {H.-B.}\ \bibnamefont
  {Chen}}, \bibinfo {author} {\bibfnamefont {J.-Y.}\ \bibnamefont {Lien}},
  \bibinfo {author} {\bibfnamefont {C.-C.}\ \bibnamefont {Hwang}}, \ and\
  \bibinfo {author} {\bibfnamefont {Y.-N.}\ \bibnamefont {Chen}},\ }\href@noop
  {} {\bibfield  {journal} {\bibinfo  {journal} {Phys. Rev. E}\ }\textbf
  {\bibinfo {volume} {89}},\ \bibinfo {pages} {042147} (\bibinfo {year}
  {2014})}\BibitemShut {NoStop}%
\bibitem [{\citenamefont {Liu}\ \emph {et~al.}(2015)\citenamefont {Liu},
  \citenamefont {Sun}, \citenamefont {Wang},\ and\ \citenamefont
  {Zhao}}]{liu15}%
  \BibitemOpen
  \bibfield  {author} {\bibinfo {author} {\bibfnamefont {J.}~\bibnamefont
  {Liu}}, \bibinfo {author} {\bibfnamefont {K.}~\bibnamefont {Sun}}, \bibinfo
  {author} {\bibfnamefont {X.}~\bibnamefont {Wang}}, \ and\ \bibinfo {author}
  {\bibfnamefont {Y.}~\bibnamefont {Zhao}},\ }\href@noop {} {\bibfield
  {journal} {\bibinfo  {journal} {Phys. Chem. Chem. Phys.}\ }\textbf {\bibinfo
  {volume} {17}},\ \bibinfo {pages} {8087} (\bibinfo {year}
  {2015})}\BibitemShut {NoStop}%
\bibitem [{\citenamefont {Chen}\ \emph {et~al.}(2015)\citenamefont {Chen},
  \citenamefont {Lambert}, \citenamefont {Cheng}, \citenamefont {Chen},\ and\
  \citenamefont {Nori}}]{chen15}%
  \BibitemOpen
  \bibfield  {author} {\bibinfo {author} {\bibfnamefont {H.-B.}\ \bibnamefont
  {Chen}}, \bibinfo {author} {\bibfnamefont {N.}~\bibnamefont {Lambert}},
  \bibinfo {author} {\bibfnamefont {Y.-C.}\ \bibnamefont {Cheng}}, \bibinfo
  {author} {\bibfnamefont {Y.-N.}\ \bibnamefont {Chen}}, \ and\ \bibinfo
  {author} {\bibfnamefont {F.}~\bibnamefont {Nori}},\ }\href@noop {} {\bibfield
   {journal} {\bibinfo  {journal} {Scientific Reports}\ }\textbf {\bibinfo
  {volume} {5}},\ \bibinfo {pages} {12753} (\bibinfo {year}
  {2015})}\BibitemShut {NoStop}%
\bibitem [{\citenamefont {Cialdi}\ \emph {et~al.}(2017)\citenamefont {Cialdi},
  \citenamefont {Rossi}, \citenamefont {Benedetti}, \citenamefont {Vacchini},
  \citenamefont {Tamascelli}, \citenamefont {Olivares},\ and\ \citenamefont
  {Paris}}]{cialdi17}%
  \BibitemOpen
  \bibfield  {author} {\bibinfo {author} {\bibfnamefont {S.}~\bibnamefont
  {Cialdi}}, \bibinfo {author} {\bibfnamefont {M.~A.~C.}\ \bibnamefont
  {Rossi}}, \bibinfo {author} {\bibfnamefont {C.}~\bibnamefont {Benedetti}},
  \bibinfo {author} {\bibfnamefont {B.}~\bibnamefont {Vacchini}}, \bibinfo
  {author} {\bibfnamefont {D.}~\bibnamefont {Tamascelli}}, \bibinfo {author}
  {\bibfnamefont {S.}~\bibnamefont {Olivares}}, \ and\ \bibinfo {author}
  {\bibfnamefont {M.~G.~A.}\ \bibnamefont {Paris}},\ }\href@noop {} {\bibfield
  {journal} {\bibinfo  {journal} {Applied Physics Letters}\ }\textbf {\bibinfo
  {volume} {110}},\ \bibinfo {pages} {081107} (\bibinfo {year}
  {2017})}\BibitemShut {NoStop}%
\bibitem [{\citenamefont {Bernardes}\ \emph {et~al.}(2015)\citenamefont
  {Bernardes}, \citenamefont {Cuevas}, \citenamefont {Orieux}, \citenamefont
  {Monken}, \citenamefont {Mataloni}, \citenamefont {Sciarrino},\ and\
  \citenamefont {Santos}}]{bernardes15}%
  \BibitemOpen
  \bibfield  {author} {\bibinfo {author} {\bibfnamefont {N.~K.}\ \bibnamefont
  {Bernardes}}, \bibinfo {author} {\bibfnamefont {A.}~\bibnamefont {Cuevas}},
  \bibinfo {author} {\bibfnamefont {A.}~\bibnamefont {Orieux}}, \bibinfo
  {author} {\bibfnamefont {C.~H.}\ \bibnamefont {Monken}}, \bibinfo {author}
  {\bibfnamefont {P.}~\bibnamefont {Mataloni}}, \bibinfo {author}
  {\bibfnamefont {F.}~\bibnamefont {Sciarrino}}, \ and\ \bibinfo {author}
  {\bibfnamefont {M.~F.}\ \bibnamefont {Santos}},\ }\href@noop {} {\bibfield
  {journal} {\bibinfo  {journal} {Scientific Reports}\ }\textbf {\bibinfo
  {volume} {5}},\ \bibinfo {pages} {17520} (\bibinfo {year}
  {2015})}\BibitemShut {NoStop}%
\bibitem [{\citenamefont {Chiuri}\ \emph {et~al.}(2012)\citenamefont {Chiuri},
  \citenamefont {Greganti}, \citenamefont {Mazzola}, \citenamefont
  {Paternostro},\ and\ \citenamefont {Mataloni}}]{chiuri12}%
  \BibitemOpen
  \bibfield  {author} {\bibinfo {author} {\bibfnamefont {A.}~\bibnamefont
  {Chiuri}}, \bibinfo {author} {\bibfnamefont {C.}~\bibnamefont {Greganti}},
  \bibinfo {author} {\bibfnamefont {L.}~\bibnamefont {Mazzola}}, \bibinfo
  {author} {\bibfnamefont {M.}~\bibnamefont {Paternostro}}, \ and\ \bibinfo
  {author} {\bibfnamefont {P.}~\bibnamefont {Mataloni}},\ }\href@noop {}
  {\bibfield  {journal} {\bibinfo  {journal} {Scientific Reports}\ }\textbf
  {\bibinfo {volume} {2}},\ \bibinfo {pages} {968} (\bibinfo {year}
  {2012})}\BibitemShut {NoStop}%
\bibitem [{\citenamefont {Liu}\ \emph {et~al.}(2011)\citenamefont {Liu},
  \citenamefont {Li}, \citenamefont {Huang}, \citenamefont {Li}, \citenamefont
  {Guo}, \citenamefont {Laine}, \citenamefont {Breuer},\ and\ \citenamefont
  {Piilo}}]{liu11}%
  \BibitemOpen
  \bibfield  {author} {\bibinfo {author} {\bibfnamefont {B.-H.}\ \bibnamefont
  {Liu}}, \bibinfo {author} {\bibfnamefont {L.}~\bibnamefont {Li}}, \bibinfo
  {author} {\bibfnamefont {Y.-F.}\ \bibnamefont {Huang}}, \bibinfo {author}
  {\bibfnamefont {C.-F.}\ \bibnamefont {Li}}, \bibinfo {author} {\bibfnamefont
  {G.-C.}\ \bibnamefont {Guo}}, \bibinfo {author} {\bibfnamefont {E.-M.}\
  \bibnamefont {Laine}}, \bibinfo {author} {\bibfnamefont {H.-P.}\ \bibnamefont
  {Breuer}}, \ and\ \bibinfo {author} {\bibfnamefont {J.}~\bibnamefont
  {Piilo}},\ }\href@noop {} {\bibfield  {journal} {\bibinfo  {journal} {Nature
  Physics}\ }\textbf {\bibinfo {volume} {7}},\ \bibinfo {pages} {931} (\bibinfo
  {year} {2011})}\BibitemShut {NoStop}%
\bibitem [{\citenamefont {{Haase}}\ \emph {et~al.}(2018)\citenamefont
  {{Haase}}, \citenamefont {{Vetter}}, \citenamefont {{Unden}}, \citenamefont
  {{Smirne}}, \citenamefont {{Rosskopf}}, \citenamefont {{Naydenov}},
  \citenamefont {{Jelezko}}, \citenamefont {{Plenio}},\ and\ \citenamefont
  {{Huelga}}}]{haase18}%
  \BibitemOpen
  \bibfield  {author} {\bibinfo {author} {\bibfnamefont {J.~F.}\ \bibnamefont
  {{Haase}}}, \bibinfo {author} {\bibfnamefont {P.~J.}\ \bibnamefont
  {{Vetter}}}, \bibinfo {author} {\bibfnamefont {T.}~\bibnamefont {{Unden}}},
  \bibinfo {author} {\bibfnamefont {A.}~\bibnamefont {{Smirne}}}, \bibinfo
  {author} {\bibfnamefont {J.}~\bibnamefont {{Rosskopf}}}, \bibinfo {author}
  {\bibfnamefont {B.}~\bibnamefont {{Naydenov}}}, \bibinfo {author}
  {\bibfnamefont {F.}~\bibnamefont {{Jelezko}}}, \bibinfo {author}
  {\bibfnamefont {M.~B.}\ \bibnamefont {{Plenio}}}, \ and\ \bibinfo {author}
  {\bibfnamefont {S.~F.}\ \bibnamefont {{Huelga}}},\ }\href@noop {} {\bibfield
  {journal} {\bibinfo  {journal} {ArXiv e-prints}\ } (\bibinfo {year}
  {2018})},\ \Eprint {http://arxiv.org/abs/1802.00819} {arXiv:1802.00819
  [quant-ph]} \BibitemShut {NoStop}%
\bibitem [{\citenamefont {{Peng}}\ \emph {et~al.}(2018)\citenamefont {{Peng}},
  \citenamefont {{Xu}}, \citenamefont {{Xu}}, \citenamefont {{Huang}},
  \citenamefont {{Wang}}, \citenamefont {{Kong}}, \citenamefont {{Rong}},
  \citenamefont {{Shi}}, \citenamefont {{Duan}},\ and\ \citenamefont
  {{Du}}}]{peng18}%
  \BibitemOpen
  \bibfield  {author} {\bibinfo {author} {\bibfnamefont {S.}~\bibnamefont
  {{Peng}}}, \bibinfo {author} {\bibfnamefont {X.}~\bibnamefont {{Xu}}},
  \bibinfo {author} {\bibfnamefont {K.}~\bibnamefont {{Xu}}}, \bibinfo {author}
  {\bibfnamefont {P.}~\bibnamefont {{Huang}}}, \bibinfo {author} {\bibfnamefont
  {P.}~\bibnamefont {{Wang}}}, \bibinfo {author} {\bibfnamefont
  {X.}~\bibnamefont {{Kong}}}, \bibinfo {author} {\bibfnamefont
  {X.}~\bibnamefont {{Rong}}}, \bibinfo {author} {\bibfnamefont
  {F.}~\bibnamefont {{Shi}}}, \bibinfo {author} {\bibfnamefont
  {C.}~\bibnamefont {{Duan}}}, \ and\ \bibinfo {author} {\bibfnamefont
  {J.}~\bibnamefont {{Du}}},\ }\href@noop {} {\bibfield  {journal} {\bibinfo
  {journal} {ArXiv e-prints}\ } (\bibinfo {year} {2018})},\ \Eprint
  {http://arxiv.org/abs/1801.04681} {arXiv:1801.04681 [quant-ph]} \BibitemShut
  {NoStop}%
\bibitem [{\citenamefont {{Wang}}\ \emph {et~al.}(2018)\citenamefont {{Wang}},
  \citenamefont {{Hou}}, \citenamefont {{Huang}}, \citenamefont {{Zhang}},
  \citenamefont {{Ouyang}}, \citenamefont {{Wang}}, \citenamefont {{Huang}},
  \citenamefont {{Zhang}}, \citenamefont {{He}}, \citenamefont {{Chang}},\ and\
  \citenamefont {{Duan}}}]{wang18}%
  \BibitemOpen
  \bibfield  {author} {\bibinfo {author} {\bibfnamefont {F.}~\bibnamefont
  {{Wang}}}, \bibinfo {author} {\bibfnamefont {P.-Y.}\ \bibnamefont {{Hou}}},
  \bibinfo {author} {\bibfnamefont {Y.-Y.}\ \bibnamefont {{Huang}}}, \bibinfo
  {author} {\bibfnamefont {W.-G.}\ \bibnamefont {{Zhang}}}, \bibinfo {author}
  {\bibfnamefont {X.-L.}\ \bibnamefont {{Ouyang}}}, \bibinfo {author}
  {\bibfnamefont {X.}~\bibnamefont {{Wang}}}, \bibinfo {author} {\bibfnamefont
  {X.-Z.}\ \bibnamefont {{Huang}}}, \bibinfo {author} {\bibfnamefont {H.-L.}\
  \bibnamefont {{Zhang}}}, \bibinfo {author} {\bibfnamefont {L.}~\bibnamefont
  {{He}}}, \bibinfo {author} {\bibfnamefont {X.-Y.}\ \bibnamefont {{Chang}}}, \
  and\ \bibinfo {author} {\bibfnamefont {L.-M.}\ \bibnamefont {{Duan}}},\
  }\href@noop {} {\bibfield  {journal} {\bibinfo  {journal} {ArXiv e-prints}\ }
  (\bibinfo {year} {2018})},\ \Eprint {http://arxiv.org/abs/1801.02729}
  {arXiv:1801.02729 [quant-ph]} \BibitemShut {NoStop}%
\bibitem [{\citenamefont {Man}\ \emph {et~al.}(2014)\citenamefont {Man},
  \citenamefont {An},\ and\ \citenamefont {Xia}}]{man14}%
  \BibitemOpen
  \bibfield  {author} {\bibinfo {author} {\bibfnamefont {Z.-X.}\ \bibnamefont
  {Man}}, \bibinfo {author} {\bibfnamefont {N.~B.}\ \bibnamefont {An}}, \ and\
  \bibinfo {author} {\bibfnamefont {Y.-J.}\ \bibnamefont {Xia}},\ }\href@noop
  {} {\bibfield  {journal} {\bibinfo  {journal} {Phys. Rev. A}\ }\textbf
  {\bibinfo {volume} {90}},\ \bibinfo {pages} {062104} (\bibinfo {year}
  {2014})}\BibitemShut {NoStop}%
\bibitem [{\citenamefont {Kutvonen}\ \emph {et~al.}(2015)\citenamefont
  {Kutvonen}, \citenamefont {Ala-Nissila},\ and\ \citenamefont
  {Pekola}}]{kutvonen15}%
  \BibitemOpen
  \bibfield  {author} {\bibinfo {author} {\bibfnamefont {A.}~\bibnamefont
  {Kutvonen}}, \bibinfo {author} {\bibfnamefont {T.}~\bibnamefont
  {Ala-Nissila}}, \ and\ \bibinfo {author} {\bibfnamefont {J.}~\bibnamefont
  {Pekola}},\ }\href@noop {} {\bibfield  {journal} {\bibinfo  {journal} {Phys.
  Rev. E}\ }\textbf {\bibinfo {volume} {92}},\ \bibinfo {pages} {012107}
  (\bibinfo {year} {2015})}\BibitemShut {NoStop}%
\bibitem [{\citenamefont {Megier}\ \emph {et~al.}(2017)\citenamefont {Megier},
  \citenamefont {Chruscinski}, \citenamefont {Piilo},\ and\ \citenamefont
  {Strunz}}]{megier2017}%
  \BibitemOpen
  \bibfield  {author} {\bibinfo {author} {\bibfnamefont {N.}~\bibnamefont
  {Megier}}, \bibinfo {author} {\bibfnamefont {D.}~\bibnamefont {Chruscinski}},
  \bibinfo {author} {\bibfnamefont {J.}~\bibnamefont {Piilo}}, \ and\ \bibinfo
  {author} {\bibfnamefont {W.~T.}\ \bibnamefont {Strunz}},\ }\href@noop {}
  {\bibfield  {journal} {\bibinfo  {journal} {Scientific Reports}\ }\textbf
  {\bibinfo {volume} {7}},\ \bibinfo {pages} {6379} (\bibinfo {year}
  {2017})}\BibitemShut {NoStop}%
\bibitem [{\citenamefont {Breuer}\ \emph {et~al.}(2018)\citenamefont {Breuer},
  \citenamefont {Amato},\ and\ \citenamefont {Vacchini}}]{breuer18}%
  \BibitemOpen
  \bibfield  {author} {\bibinfo {author} {\bibfnamefont {H.-P.}\ \bibnamefont
  {Breuer}}, \bibinfo {author} {\bibfnamefont {G.}~\bibnamefont {Amato}}, \
  and\ \bibinfo {author} {\bibfnamefont {B.}~\bibnamefont {Vacchini}},\
  }\href@noop {} {\bibfield  {journal} {\bibinfo  {journal} {New Journal of
  Physics}\ } (\bibinfo {year} {2018})}\BibitemShut {NoStop}%
\bibitem [{\citenamefont {Bellomo}\ \emph {et~al.}(2007)\citenamefont
  {Bellomo}, \citenamefont {Lo~Franco},\ and\ \citenamefont
  {Compagno}}]{bellomo07}%
  \BibitemOpen
  \bibfield  {author} {\bibinfo {author} {\bibfnamefont {B.}~\bibnamefont
  {Bellomo}}, \bibinfo {author} {\bibfnamefont {R.}~\bibnamefont {Lo~Franco}},
  \ and\ \bibinfo {author} {\bibfnamefont {G.}~\bibnamefont {Compagno}},\
  }\href@noop {} {\bibfield  {journal} {\bibinfo  {journal} {Phys. Rev. Lett.}\
  }\textbf {\bibinfo {volume} {99}},\ \bibinfo {pages} {160502} (\bibinfo
  {year} {2007})}\BibitemShut {NoStop}%
\bibitem [{\citenamefont {Bellomo}\ \emph {et~al.}(2008)\citenamefont
  {Bellomo}, \citenamefont {Lo~Franco},\ and\ \citenamefont
  {Compagno}}]{bellomo08}%
  \BibitemOpen
  \bibfield  {author} {\bibinfo {author} {\bibfnamefont {B.}~\bibnamefont
  {Bellomo}}, \bibinfo {author} {\bibfnamefont {R.}~\bibnamefont {Lo~Franco}},
  \ and\ \bibinfo {author} {\bibfnamefont {G.}~\bibnamefont {Compagno}},\
  }\href@noop {} {\bibfield  {journal} {\bibinfo  {journal} {Phys. Rev. A}\
  }\textbf {\bibinfo {volume} {77}},\ \bibinfo {pages} {032342} (\bibinfo
  {year} {2008})}\BibitemShut {NoStop}%
\bibitem [{\citenamefont {Abel}\ and\ \citenamefont
  {Marquardt}(2008)}]{Abel2008}%
  \BibitemOpen
  \bibfield  {author} {\bibinfo {author} {\bibfnamefont {B.}~\bibnamefont
  {Abel}}\ and\ \bibinfo {author} {\bibfnamefont {F.}~\bibnamefont
  {Marquardt}},\ }\href@noop {} {\bibfield  {journal} {\bibinfo  {journal}
  {Phys. Rev. B}\ }\textbf {\bibinfo {volume} {78}},\ \bibinfo {pages}
  {201302R} (\bibinfo {year} {2008})}\BibitemShut {NoStop}%
\bibitem [{\citenamefont {Rossi}\ and\ \citenamefont
  {Paris}(2016{\natexlab{b}})}]{Rossi2016}%
  \BibitemOpen
  \bibfield  {author} {\bibinfo {author} {\bibfnamefont {M.~A.~C.}\
  \bibnamefont {Rossi}}\ and\ \bibinfo {author} {\bibfnamefont {M.~G.~A.}\
  \bibnamefont {Paris}},\ }\href@noop {} {\bibfield  {journal} {\bibinfo
  {journal} {J. Chem. Phys.}\ }\textbf {\bibinfo {volume} {144}},\ \bibinfo
  {pages} {024113} (\bibinfo {year} {2016}{\natexlab{b}})}\BibitemShut
  {NoStop}%
\bibitem [{\citenamefont {Gehlen}\ \emph {et~al.}(1994)\citenamefont {Gehlen},
  \citenamefont {Marchi},\ and\ \citenamefont {Chandler}}]{Gehlen1994}%
  \BibitemOpen
  \bibfield  {author} {\bibinfo {author} {\bibfnamefont {J.~H.}\ \bibnamefont
  {Gehlen}}, \bibinfo {author} {\bibfnamefont {M.}~\bibnamefont {Marchi}}, \
  and\ \bibinfo {author} {\bibfnamefont {D.}~\bibnamefont {Chandler}},\
  }\href@noop {} {\bibfield  {journal} {\bibinfo  {journal} {Science}\ }\textbf
  {\bibinfo {volume} {263}},\ \bibinfo {pages} {5067} (\bibinfo {year}
  {1994})}\BibitemShut {NoStop}%
\bibitem [{\citenamefont {Petrov}\ \emph {et~al.}(1994)\citenamefont {Petrov},
  \citenamefont {Teslenko},\ and\ \citenamefont {Goychuk}}]{Petrov1994}%
  \BibitemOpen
  \bibfield  {author} {\bibinfo {author} {\bibfnamefont {E.~G.}\ \bibnamefont
  {Petrov}}, \bibinfo {author} {\bibfnamefont {V.~I.}\ \bibnamefont
  {Teslenko}}, \ and\ \bibinfo {author} {\bibfnamefont {I.~A.}\ \bibnamefont
  {Goychuk}},\ }\href@noop {} {\bibfield  {journal} {\bibinfo  {journal} {Phys.
  Rev. E}\ }\textbf {\bibinfo {volume} {49}},\ \bibinfo {pages} {3894}
  (\bibinfo {year} {1994})}\BibitemShut {NoStop}%
\bibitem [{\citenamefont {Goychuk}\ \emph
  {et~al.}(1995{\natexlab{a}})\citenamefont {Goychuk}, \citenamefont {Petrov},\
  and\ \citenamefont {May}}]{Goychuk1995}%
  \BibitemOpen
  \bibfield  {author} {\bibinfo {author} {\bibfnamefont {I.~A.}\ \bibnamefont
  {Goychuk}}, \bibinfo {author} {\bibfnamefont {E.~G.}\ \bibnamefont {Petrov}},
  \ and\ \bibinfo {author} {\bibfnamefont {V.}~\bibnamefont {May}},\
  }\href@noop {} {\bibfield  {journal} {\bibinfo  {journal} {Phys. Rev. E}\
  }\textbf {\bibinfo {volume} {52}},\ \bibinfo {pages} {3} (\bibinfo {year}
  {1995}{\natexlab{a}})}\BibitemShut {NoStop}%
\bibitem [{\citenamefont {Goychuk}\ \emph
  {et~al.}(1997{\natexlab{a}})\citenamefont {Goychuk}, \citenamefont {Petrov},\
  and\ \citenamefont {May}}]{Combined1997}%
  \BibitemOpen
  \bibfield  {author} {\bibinfo {author} {\bibfnamefont {I.~A.}\ \bibnamefont
  {Goychuk}}, \bibinfo {author} {\bibfnamefont {E.~G.}\ \bibnamefont {Petrov}},
  \ and\ \bibinfo {author} {\bibfnamefont {V.}~\bibnamefont {May}},\
  }\href@noop {} {\bibfield  {journal} {\bibinfo  {journal} {Phys. Rev. E}\
  }\textbf {\bibinfo {volume} {56}},\ \bibinfo {pages} {2} (\bibinfo {year}
  {1997}{\natexlab{a}})}\BibitemShut {NoStop}%
\bibitem [{\citenamefont {May}\ and\ \citenamefont {K\"{u}hn}(2003)}]{May2003}%
  \BibitemOpen
  \bibfield  {author} {\bibinfo {author} {\bibfnamefont {V.}~\bibnamefont
  {May}}\ and\ \bibinfo {author} {\bibfnamefont {O.}~\bibnamefont {K\"{u}hn}},\
  }\href@noop {} {\bibfield  {journal} {\bibinfo  {journal} {\emph{Charge and
  Energy Transfer Dynamics in Molecular Systems} (Wiley-VCH, Berlin, 2003)}\ }
  (\bibinfo {year} {2003})}\BibitemShut {NoStop}%
\bibitem [{\citenamefont {Albinsson}\ and\ \citenamefont
  {Maertensson}(2008)}]{Albinsson2008}%
  \BibitemOpen
  \bibfield  {author} {\bibinfo {author} {\bibfnamefont {B.}~\bibnamefont
  {Albinsson}}\ and\ \bibinfo {author} {\bibfnamefont {J.}~\bibnamefont
  {Maertensson}},\ }\href@noop {} {\bibfield  {journal} {\bibinfo  {journal}
  {J. Photochem. Photobio., A}\ }\textbf {\bibinfo {volume} {9}},\ \bibinfo
  {pages} {138 } (\bibinfo {year} {2008})}\BibitemShut {NoStop}%
\bibitem [{\citenamefont {Tang}(1993)}]{Tang1993}%
  \BibitemOpen
  \bibfield  {author} {\bibinfo {author} {\bibfnamefont {J.}~\bibnamefont
  {Tang}},\ }\href@noop {} {\bibfield  {journal} {\bibinfo  {journal} {J. Chem.
  Phys.}\ }\textbf {\bibinfo {volume} {98}},\ \bibinfo {pages} {15} (\bibinfo
  {year} {1993})}\BibitemShut {NoStop}%
\bibitem [{\citenamefont {Goychuk}\ \emph
  {et~al.}(1995{\natexlab{b}})\citenamefont {Goychuk}, \citenamefont {Petrov},\
  and\ \citenamefont {May}}]{BridgeAssist1995}%
  \BibitemOpen
  \bibfield  {author} {\bibinfo {author} {\bibfnamefont {I.~A.}\ \bibnamefont
  {Goychuk}}, \bibinfo {author} {\bibfnamefont {E.~G.}\ \bibnamefont {Petrov}},
  \ and\ \bibinfo {author} {\bibfnamefont {V.}~\bibnamefont {May}},\
  }\href@noop {} {\bibfield  {journal} {\bibinfo  {journal} {J. Chem. Phys.}\
  }\textbf {\bibinfo {volume} {103}},\ \bibinfo {pages} {12} (\bibinfo {year}
  {1995}{\natexlab{b}})}\BibitemShut {NoStop}%
\bibitem [{\citenamefont {Goychuk}\ \emph
  {et~al.}(1995{\natexlab{c}})\citenamefont {Goychuk}, \citenamefont {Petrov},\
  and\ \citenamefont {May}}]{Dissipative1995}%
  \BibitemOpen
  \bibfield  {author} {\bibinfo {author} {\bibfnamefont {I.~A.}\ \bibnamefont
  {Goychuk}}, \bibinfo {author} {\bibfnamefont {E.~G.}\ \bibnamefont {Petrov}},
  \ and\ \bibinfo {author} {\bibfnamefont {V.}~\bibnamefont {May}},\
  }\href@noop {} {\bibfield  {journal} {\bibinfo  {journal} {Phys. Rev. E}\
  }\textbf {\bibinfo {volume} {51}},\ \bibinfo {pages} {4} (\bibinfo {year}
  {1995}{\natexlab{c}})}\BibitemShut {NoStop}%
\bibitem [{\citenamefont {Goychuk}\ \emph
  {et~al.}(1997{\natexlab{b}})\citenamefont {Goychuk}, \citenamefont {Petrov},\
  and\ \citenamefont {May}}]{Control1997}%
  \BibitemOpen
  \bibfield  {author} {\bibinfo {author} {\bibfnamefont {I.~A.}\ \bibnamefont
  {Goychuk}}, \bibinfo {author} {\bibfnamefont {E.~G.}\ \bibnamefont {Petrov}},
  \ and\ \bibinfo {author} {\bibfnamefont {V.}~\bibnamefont {May}},\
  }\href@noop {} {\bibfield  {journal} {\bibinfo  {journal} {J. Chem. Phys.}\
  }\textbf {\bibinfo {volume} {106}},\ \bibinfo {pages} {11} (\bibinfo {year}
  {1997}{\natexlab{b}})}\BibitemShut {NoStop}%
\bibitem [{\citenamefont {Iwaniszewski}(2000)}]{Jan2000}%
  \BibitemOpen
  \bibfield  {author} {\bibinfo {author} {\bibfnamefont {J.}~\bibnamefont
  {Iwaniszewski}},\ }\href@noop {} {\bibfield  {journal} {\bibinfo  {journal}
  {Phys. Rev. E}\ }\textbf {\bibinfo {volume} {61}},\ \bibinfo {pages} {4890}
  (\bibinfo {year} {2000})}\BibitemShut {NoStop}%
\bibitem [{\citenamefont {Goychuk}(1995)}]{Kinetic1995}%
  \BibitemOpen
  \bibfield  {author} {\bibinfo {author} {\bibfnamefont {I.~A.}\ \bibnamefont
  {Goychuk}},\ }\href@noop {} {\bibfield  {journal} {\bibinfo  {journal} {Phys.
  Rev. E}\ }\textbf {\bibinfo {volume} {51}},\ \bibinfo {pages} {6} (\bibinfo
  {year} {1995})}\BibitemShut {NoStop}%
\bibitem [{\citenamefont {Goychuk}\ and\ \citenamefont
  {Hanggi}(2005)}]{goychuk05}%
  \BibitemOpen
  \bibfield  {author} {\bibinfo {author} {\bibfnamefont {I.}~\bibnamefont
  {Goychuk}}\ and\ \bibinfo {author} {\bibfnamefont {P.}~\bibnamefont
  {Hanggi}},\ }\href@noop {} {\bibfield  {journal} {\bibinfo  {journal}
  {Advances in Physics}\ }\textbf {\bibinfo {volume} {54}},\ \bibinfo {pages}
  {525} (\bibinfo {year} {2005})}\BibitemShut {NoStop}%
\bibitem [{\citenamefont {Mazzola}\ \emph {et~al.}(2010)\citenamefont
  {Mazzola}, \citenamefont {Laine}, \citenamefont {Breuer}, \citenamefont
  {Maniscalco},\ and\ \citenamefont {Piilo}}]{mazzola10}%
  \BibitemOpen
  \bibfield  {author} {\bibinfo {author} {\bibfnamefont {L.}~\bibnamefont
  {Mazzola}}, \bibinfo {author} {\bibfnamefont {E.-M.}\ \bibnamefont {Laine}},
  \bibinfo {author} {\bibfnamefont {H.-P.}\ \bibnamefont {Breuer}}, \bibinfo
  {author} {\bibfnamefont {S.}~\bibnamefont {Maniscalco}}, \ and\ \bibinfo
  {author} {\bibfnamefont {J.}~\bibnamefont {Piilo}},\ }\href@noop {}
  {\bibfield  {journal} {\bibinfo  {journal} {Phys. Rev. A}\ }\textbf {\bibinfo
  {volume} {81}},\ \bibinfo {pages} {062120} (\bibinfo {year}
  {2010})}\BibitemShut {NoStop}%
\bibitem [{\citenamefont {Onuchic}\ \emph {et~al.}(1986)\citenamefont
  {Onuchic}, \citenamefont {Beratan},\ and\ \citenamefont
  {Hopfield}}]{Hopfield1986}%
  \BibitemOpen
  \bibfield  {author} {\bibinfo {author} {\bibfnamefont {J.}~\bibnamefont
  {Onuchic}}, \bibinfo {author} {\bibfnamefont {D.}~\bibnamefont {Beratan}}, \
  and\ \bibinfo {author} {\bibfnamefont {J.}~\bibnamefont {Hopfield}},\
  }\href@noop {} {\bibfield  {journal} {\bibinfo  {journal} {J. Phys. Chem.}\
  }\textbf {\bibinfo {volume} {90}},\ \bibinfo {pages} {3707} (\bibinfo {year}
  {1986})}\BibitemShut {NoStop}%
\bibitem [{\citenamefont {Garg}\ \emph {et~al.}(1985)\citenamefont {Garg},
  \citenamefont {Onuchic},\ and\ \citenamefont {Ambegaokar}}]{garg1985}%
  \BibitemOpen
  \bibfield  {author} {\bibinfo {author} {\bibfnamefont {A.}~\bibnamefont
  {Garg}}, \bibinfo {author} {\bibfnamefont {J.~N.}\ \bibnamefont {Onuchic}}, \
  and\ \bibinfo {author} {\bibfnamefont {V.}~\bibnamefont {Ambegaokar}},\
  }\href@noop {} {\bibfield  {journal} {\bibinfo  {journal} {J. Chem. Phys.}\
  }\textbf {\bibinfo {volume} {83}},\ \bibinfo {pages} {4491} (\bibinfo {year}
  {1985})}\BibitemShut {NoStop}%
\bibitem [{\citenamefont {Nazir}\ and\ \citenamefont
  {McCutcheon}(2016)}]{Nazir2016}%
  \BibitemOpen
  \bibfield  {author} {\bibinfo {author} {\bibfnamefont {A.}~\bibnamefont
  {Nazir}}\ and\ \bibinfo {author} {\bibfnamefont {D.~P.~S.}\ \bibnamefont
  {McCutcheon}},\ }\href@noop {} {\bibfield  {journal} {\bibinfo  {journal} {J.
  Phys.: Condens. Matter}\ }\textbf {\bibinfo {volume} {28}},\ \bibinfo {pages}
  {10} (\bibinfo {year} {2016})}\BibitemShut {NoStop}%
\bibitem [{\citenamefont {Weiss}(1998)}]{weiss98}%
  \BibitemOpen
  \bibfield  {author} {\bibinfo {author} {\bibfnamefont {U.}~\bibnamefont
  {Weiss}},\ }\href@noop {} {\emph {\bibinfo {title} {Quantum Dissipative
  Systems}}}\ (\bibinfo  {publisher} {World Scientific, Singapore},\ \bibinfo
  {year} {1998})\BibitemShut {NoStop}%
\bibitem [{\citenamefont {Bourret}\ \emph {et~al.}(1973)\citenamefont
  {Bourret}, \citenamefont {Frisch},\ and\ \citenamefont
  {Pouquet}}]{Bourret73}%
  \BibitemOpen
  \bibfield  {author} {\bibinfo {author} {\bibfnamefont {R.~C.}\ \bibnamefont
  {Bourret}}, \bibinfo {author} {\bibfnamefont {U.}~\bibnamefont {Frisch}}, \
  and\ \bibinfo {author} {\bibfnamefont {A.}~\bibnamefont {Pouquet}},\
  }\href@noop {} {\bibfield  {journal} {\bibinfo  {journal} {Physica}\ }\textbf
  {\bibinfo {volume} {65}},\ \bibinfo {pages} {303} (\bibinfo {year}
  {1973})}\BibitemShut {NoStop}%
\bibitem [{\citenamefont {Shapiro}\ and\ \citenamefont
  {Loginov}(1978)}]{Shapiro78}%
  \BibitemOpen
  \bibfield  {author} {\bibinfo {author} {\bibfnamefont {V.~E.}\ \bibnamefont
  {Shapiro}}\ and\ \bibinfo {author} {\bibfnamefont {V.~M.}\ \bibnamefont
  {Loginov}},\ }\href@noop {} {\bibfield  {journal} {\bibinfo  {journal}
  {Physica A}\ }\textbf {\bibinfo {volume} {91}},\ \bibinfo {pages} {563}
  (\bibinfo {year} {1978})}\BibitemShut {NoStop}%
\bibitem [{\citenamefont {Magazzu}\ \emph {et~al.}(2017)\citenamefont
  {Magazzu}, \citenamefont {H\"anggi}, \citenamefont {Spagnolo},\ and\
  \citenamefont {Valenti}}]{Luca2017}%
  \BibitemOpen
  \bibfield  {author} {\bibinfo {author} {\bibfnamefont {L.}~\bibnamefont
  {Magazzu}}, \bibinfo {author} {\bibfnamefont {P.}~\bibnamefont {H\"anggi}},
  \bibinfo {author} {\bibfnamefont {B.}~\bibnamefont {Spagnolo}}, \ and\
  \bibinfo {author} {\bibfnamefont {D.}~\bibnamefont {Valenti}},\ }\href
  {\doibase 10.1103/PhysRevE.95.042104} {\bibfield  {journal} {\bibinfo
  {journal} {Phys. Rev. E}\ }\textbf {\bibinfo {volume} {95}},\ \bibinfo
  {pages} {042104} (\bibinfo {year} {2017})}\BibitemShut {NoStop}%
\bibitem [{\citenamefont {Wissmann}\ \emph {et~al.}(2012)\citenamefont
  {Wissmann}, \citenamefont {Karlsson}, \citenamefont {Laine}, \citenamefont
  {Piilo},\ and\ \citenamefont {Breuer}}]{wissmann12}%
  \BibitemOpen
  \bibfield  {author} {\bibinfo {author} {\bibfnamefont {S.}~\bibnamefont
  {Wissmann}}, \bibinfo {author} {\bibfnamefont {A.}~\bibnamefont {Karlsson}},
  \bibinfo {author} {\bibfnamefont {E.-M.}\ \bibnamefont {Laine}}, \bibinfo
  {author} {\bibfnamefont {J.}~\bibnamefont {Piilo}}, \ and\ \bibinfo {author}
  {\bibfnamefont {H.-P.}\ \bibnamefont {Breuer}},\ }\href@noop {} {\bibfield
  {journal} {\bibinfo  {journal} {Phys. Rev. A}\ }\textbf {\bibinfo {volume}
  {86}},\ \bibinfo {pages} {062108} (\bibinfo {year} {2012})}\BibitemShut
  {NoStop}%
\bibitem [{\citenamefont {Rivas}(2017)}]{rivas17}%
  \BibitemOpen
  \bibfield  {author} {\bibinfo {author} {\bibfnamefont {A.}~\bibnamefont
  {Rivas}},\ }\href@noop {} {\bibfield  {journal} {\bibinfo  {journal} {Phys.
  Rev. A}\ }\textbf {\bibinfo {volume} {95}},\ \bibinfo {pages} {042104}
  (\bibinfo {year} {2017})}\BibitemShut {NoStop}%
\end{thebibliography}
%

\end{document}